\documentclass[aps,prd,nofootinbib,preprint]{revtex4}
\usepackage{graphicx}

\begin{document}
\title{Radiative Corrections on the $B\to P$ Form Factors
with Chiral Current in the Light-Cone Sum Rules}
\date{\today}
\author{Xing-Gang Wu$^{1}$ \footnote{email: wuxg@cqu.edu.cn} and
Tao Huang$^{2}$\footnote{email: huangtao@mail.ihep.ac.cn}}
\address{$^1$Department of Physics, Chongqing University, Chongqing 400030,
P.R. China\\ $^2$Institute of High Energy Physics and Theoretical
Physics Center for Science Facilities, Chinese Academy of Sciences,
P.O.Box 918(4), Beijing 100049, P.R. China}

\begin{abstract}
Based on the approach of the vector form factor $F^{+}_{B\to\pi,
K}(q^2)$ in our previous papers, we extend the calculation of the
radiative corrections to the $B\to P$ ($P$ stands $\pi$, $K$ and all
light pseudoscalar mesons) scalar and tensor form factors
$F^{0,T}_{B\to P}(q^2)$ with chiral current in the light-cone sum
rules (LCSRs). The most uncertain twist-3 contributions to the $B\to
P$ form factors can be naturally eliminated through a properly
designed correlator. We present the next-leading-order formulae of
$F^{+,0,T}_{B\to P}(q^2)$ with the $b$-quark pole mass that is
universal. It has been shown that our results are simpler and less
uncertain under the same parameter regions since we only need to
calculate the next leading order on the twist-2 part from the
obtained LCSR. Second, we obtain
$f^{+,0}_{B\to\pi}(0)=0.260^{+0.059}_{-0.040}$,
$f^{T}_{B\to\pi}(0)=0.276^{+0.052}_{-0.039}$, $f^{+,0}_{B\to
K}(0)=0.334^{+0.094}_{-0.069}$ and $f^{T}_{B\to K}(0)
=0.379^{+0.092}_{-0.077}$ at $q^2=0$ and the $SU_f(3)$-breaking
effects are discussed too. \\

\noindent {\bf PACS numbers:} 14.40.Aq, 12.38.Bx, 13.20.He, 11.55.Hx

\noindent {\bf Keywords:} B-physics, NLO Calculations, QCD LCSR,
Chiral Current

\end{abstract}

\maketitle

\section{Introduction}

The form factors of heavy-to-light transitions at large and
intermediate energies are among the most important applications of
QCD light-cone sum rule (LCSR), since the validity of the LCSR
approach is restricted to the large meson energy
($E_P>>\Lambda_{QCD}$) via the relation $q^2=m_B^2-2m_B E_P$. In
literature there are several approaches to calculate the $B\to {\rm
light\; meson}$ transition form factors in addition to the QCD LCSR
approach, such as the lattice QCD technique and the perturbative QCD
(PQCD) approach. These approaches are complementary to each other,
since they are adaptable in different energy regions, and by
combining the results from these three methods, one may obtain a
full understanding of the $B\to {\rm light \; meson}$ transition
form factors in its whole physical region \cite{hwbpi,hqw,wk1,wk2}.
Since the LCSR is restricted to small and moderate $q^2$, a better
LCSR shall present a better connection to both the PQCD and the
lattice QCD results, and then a better understanding of these form
factors.

How to ``design" a proper correlator for these heavy-to-light form
factors is a tricky problem. If the correlator is chosen properly,
one can simplify the LCSR greatly. As for the $B\to$ light
pseudoscalar mesons, the commonly adopted correlators are usually
defined as
\begin{equation}
\Pi^{\pm}_\mu (p,q) =i\int d^4xe^{iq\cdot x}\langle P(p)|T \{
\overline{q}(x)\gamma _\mu b(x), \overline{b}(0) i m_b \gamma_5
q'(0)\}|0\rangle \label{cc1}
\end{equation}
and
\begin{equation}
\Pi^T_\mu (p,q)=i\int d^4xe^{iq\cdot x}\langle P(p)|T \{
\overline{q}(x)i\sigma_{\mu\nu}q^{\nu}b(x), \overline{b}(0)i m_b
\gamma_{5} q'(0)\}|0\rangle ,\label{cc2}
\end{equation}
where $q(x)$ and $q'(0)$ stand for the light quark fields that form
the pseudo-scalar mesons. By taking such conventional correlation
functions, it has been found that the main uncertainties in
estimation of the $B\to P$ form factors come from the different
twist structures of pion/kaon wave functions, and most importantly,
the twist-2 and twist-3 contributions should be treated on the equal
footing \cite{pball2,pball3,melic}. Thus one has to calculate both
the twist-2 and twist-3 contributions up to one-loop accuracy in
order to obtain a consistent one-loop estimation of the form
factors.

On the other hand, by taking proper chiral currents into the
correlator, one can directly eliminate the most uncertain twist-3
terms, and then only needs to calculate the twist-2 contribution to
next-to-leading order (NLO) accuracy
\cite{huangbpi1,huangbpi2,hlwz}. At the present, the vector form
factors $f^{+}_{B\to\pi,K}(q^2)$ has been calculated with the chiral
current in the LCSR up to NLO \cite{huangbpi2,wk1}. It can be found
that the scalar and penguin form factors $f^{0,T}_{B\to\pi,K}(q^2)$
shall be important in due cases, e.g. the penguin form factors shall
give sizable contributions to $B\to P l^{+}l^{-}$ or $B\to
K^*\gamma$ \cite{huangbk}. So it is interesting to extend the
previous study to all the $B\to P$ ($P$ stands $\pi$, $K$ and all
light pseudoscalar mesons) transition form factors $f^{+,0,T}_{B\to
P}(q^2)$ with the chiral current in the LCSR up to one-loop
accuracy. Furthermore, it maybe also interesting to know to what
degree the different choices of correlator shall affect the final
LCSRs, which is another purpose of present paper.

The paper is organized as follows. In Sec. II, we present the
calculation technology to obtain the LCSRs for the $B\to P$
transition form factors $f^{+,0,T}_{B\to P}(q^2)$ with chiral
currents, where the $SU_f(3)$-breaking effects for the kaonic case
will be explained in due places. Numerical results and discussions
are presented in Sec. III, where the uncertainties of form factors
under the present LCSRs shall be discussed. The comparison with
other approaches will be presented in Sec.IV. Sec.V is reserved for
a summary.

\section{calculation technology for the $B\to P$ transition
form factors with proper chiral currents}

\subsection{A definition of $f^{+,0,T}_{B\to P}$}

Based on the previous calculation about the transition form factor
$B\to \pi/K$, we present the formulae for the $B\to P$ transition
form factors for generality such that these formulae can also be
conveniently extended for other light pseudo-scalar form factors
like $B\to\eta$ and $B\to\eta'$ form factors. With default, we adopt
the chiral limit $p^2_P=m_P^2=0$, but point out how to include the
$SU_f(3)$-breaking effects for the $B\to K$ form factors in due
places, i.e. the dominant $SU_f(3)$-breaking effects will be
discussed with the newly obtained $K$ meson distribution amplitudes
\cite{ballmoments}. The hadronic matrix elements for the $B\to P$
transition form factors are parameterized as
\begin{eqnarray}
\langle P(p_{P}) | \bar q^{\prime} \gamma^{\mu}  b| \bar
B(p_B)\rangle &=& f_{B\to P}^{+}(q^2)\left(P^{\mu}-\frac{P\cdot
q}{q^2}q^{\mu} \right)+f_{B\to P}^{0}(q^2) \frac{P\cdot q}{q^2} q_{\mu} \nonumber\\
&=& 2f_{B\to P}^{+}(q^2)p_P^{\mu}+[f_{B\to P}^{+}(q^2)+f_{B\to P}^{-}(q^2)] q_{\mu}\ ,\nonumber\\
\langle P(p_{P} )| \bar q^{\prime} i\sigma_{\mu\nu} q^{\nu}b| \bar B
(p_{B})\rangle &=& {f_{B\to P}^{T}(q^2)\over
m_{B}+m_{P}}\left[P\cdot q\, q_{\mu}-q^{2}P_{\mu}\right]
\label{bpff}
\end{eqnarray}
with $P$ representing the pseudoscalar, $P_{\mu}=(p_B+p_P)_{\mu}$,
$q_{\mu}=(p_B-p_P)_{\mu}$, and $f^{+}_{B\to P}(q^{2})$, $f^{0}_{B\to
P}(q^{2})$, $f^{T}_{B\to P}(q^{2})$ stand for the vector, scalar and
tensor form factors respectively. It can be found that the scalar
form factor $f^{0}_{B\to P}(q^2)$ satisfies the following relation:
\begin{equation}\label{relation0p}
f_{B\to P}^{0}(q^2)=f_{B\to P}^{+}(q^2) + \frac{q^2}{m_B^2-m_P^2}
f_{B\to P}^{-}(q^2) .
\end{equation}

As for the LCSR calculation, different to the conventional choice of
the correlation functions as shown in Eqs.(\ref{cc1},\ref{cc2}), we
choose the following chiral currents in the correlation functions,
\begin{eqnarray}
\Pi^{\pm}_\mu (p,q) && =i\int d^4xe^{iq\cdot x}\langle P(p)|T \{
\overline{q}(x)\gamma _\mu (1+\gamma _5)b(x), \overline{b}(0)
i m_b(1+\gamma _5)q'(0)\}|0\rangle , \nonumber\\
&& =\Pi^{+}(q^2,(p+q)^2)p_\mu +\Pi^{-}(q^2,(p+q)^2)q_\mu ,\\
\Pi^T_\mu (p,q) && =i\int d^4xe^{iq\cdot x}\langle P(p)|T \{
\overline{q}(x)i\sigma_{\mu\nu}q^{\nu} (1+\gamma _5)b(x),
\overline{b}(0)i m_b(1-\gamma _5)q'(0)\}|0\rangle , \nonumber\\
&& =\Pi^{T}(q^2,(p+q)^2)\left[(P\cdot q)q_{\mu}-q^{2}P_\mu\right] ,
\end{eqnarray}
where $P=p+2q$.

We calculate the form factors $f^{+,0,T}_{B\to P}(q^2)$ following
the same calculation technology as described in
Refs.\cite{huangbpi2,wk1}, where the vector form factors $f_{B\to
\pi, K}^{+}(q^2)$ have been calculated. For such purpose, we first
give a simple extension to $f_{B\to P}^{+}(q^2)$ in the large
space-like momentum regions $(p+q)^2-m_b^2\ll 0$ and $q^2\ll m_b^2$
for the momentum transfer, which correspond to the small light-cone
distance $x^2\approx 0$ and are required by the validity of OPE. And
then, we present the newly obtained results for the scalar and
tensor form factors.

\subsection{A simple extension to $f^{+}_{B\to P}$ within LCSR}

The vacuum-to-meson matrix elements in terms of the pseudo-scalar's
LC DAs of different twist can be expanded by contracting the
$b$-quark fields with the help of the full $b$-quark propagator
within the background field:
\begin{eqnarray}
\langle0|T{b(x)\bar{b}(0)}|0\rangle && =i\int \frac{d^4k}{(2\pi
)^4}e^{-ikx}\frac{\slash\!\!\!{k}+m}{k^2-m_b^2}-ig_s\int
\frac{d^4k}{(2\pi
)^4}e^{-ikx}\cdot \nonumber \\
&& \int_0^1 dv \left[\frac {1}{2}\frac{m+\slash\!\!\!{k}}
{(k^2-m_b^2)^2}G^{\mu \nu }(vx)\sigma _{\mu \nu } -\frac
1{k^2-m_b^2}v x_\mu G^{\mu \nu }(vx)\gamma _\nu \right],
\end{eqnarray}
where only the free propagator and the one-gluon terms are retained,
$G_{\mu \nu }$ stands for the background gluonic field strength, and
$g_s$ denotes the strong coupling constant. The invariant amplitudes
$\Pi^{+}$ can be obtained by substituting the $b$-quark propagator
and the corresponding LC wave functions, and completing the
integrations over $x$ and $k$.

The OPE results for the invariant amplitudes $\Pi^{+}$ can be
represented as a sum of LO and NLO parts:
\begin{eqnarray}
\Pi^{+}(q^2,(p+q)^2)&=& \Pi^{+}_0(q^2,(p+q)^2)+\frac{\alpha_sC_F}
{4\pi}\Pi^{+}_1(q^2,(p+q)^2), \label{corr}
\end{eqnarray}
where $\Pi^{+}_0(q^2,(p+q)^2)$ and $\Pi^{+}_1(q^2,(p+q)^2)$ stands
for the LO and the NLO contributions respectively. As for the LO
invariant amplitude, we obtain:
\begin{eqnarray}
\Pi^{+}_{0}(q^2,(p+q)^2) &=& 2f_P m^2_b \left[ \int_{0}^{1}
\frac{du}{u} \frac{\varphi _{P}(u)}{\Delta}-\int_{0}^{1}
\frac{du}{u^3} \frac{m_b^2}{2\Delta^3}\phi _{4P}(u)+\int_{0}^{1}
\frac{du}{u\Delta^2}G_{4P}(u) \nonumber  \right.\\
&&\left. + \int_0^1 dv \int D\alpha_i \frac {2 \Psi_{4P}(\alpha_i)
+2\tilde{\Psi}_{4P}(\alpha_i)
-\Phi_{4P}(\alpha_i)-\tilde{\Phi}_{4P}(\alpha_i)}
{\Delta^2(\alpha_1+v\alpha_3)^2}\right], \label{loa}
\end{eqnarray}
where the parameters are defined as: $\Delta=s-(p+q)^2$
($s=[q^2+(m_b^2-q^2)/u]$), $G_{4P}(u)=-\int_0^u dv\psi_{4P}(v)$ and
$D\alpha_i=d\alpha_1 d\alpha_2 d\alpha_3 \delta(1-\alpha_1-\alpha_2
-\alpha_3)$. Here $\varphi_P$ is the twist-2 LC wave function, and
$\phi_P(u)$, $\psi_{4P}(u)$, $\Psi_{4P}(\alpha_i)$,
$\tilde\Psi_{4P}(\alpha_i)$, $\Phi_{4P}(\alpha_i)$ and
$\tilde\Phi_{4P}(\alpha_i)$ are twist-4 LC wave functions defined in
a same way as the pionic case that have been defined in
Ref.\cite{ballmoments}, whose explicit forms are put in the APPENDIX
A. It is found that only the twist-2 and twist-4 contributions are
contained in the above expressions, and the twist-3 terms are
rightly eliminated by taking the present adopted chiral currents
within the correlators.

Since the most uncertain twist-3 contributions are eliminated and
the twist-4 contribution itself is quite small, so we only need to
consider the NLO correction to the twist-2 terms. The NLO invariant
amplitude $\Pi^{+}_1$ for the twist-2 contribution can be written in
the following factorized form:
\begin{equation}
\Pi^{+}_1(q^2,(p+q)^2) = -f_P \int_0^1 du
T^{+}_1(q^2,(p+q)^2,u)\varphi_P(u),
\end{equation}
where by taking $m_b$ to be the b-quark pole mass, the NLO hard
scattering amplitudes $T^{+}_1$ can be written as
\begin{eqnarray}
T^+_1(r_1,r_2,u)&=&\frac{6}{1-\rho}\left(2-\ln\frac{m_b^{2}}{\mu^2}\right)
-\frac{4}{1-\rho} \left[2G(\rho)-G(r_1)-G(r_2)\right]\nonumber \\
&+& \frac{4}{(r_1-r_2)^2} \left\{\frac{1-r_2}{u}[ G(\rho)-G(r_1)]+
\frac{1-r_1}{\bar{u}}\left[ G(\rho)-G(r_2)\right] \right\}  \nonumber\\
&+& 2\frac{\rho+(1-\rho)\ln\left(1-\rho\right)}{\rho^2}
-\frac{4}{1-\rho} \frac{(1-r_2)\ln\left(1-r_2\right)}{r_2}\nonumber \\
&-& \frac{4}{\rho-r_2}\left[ \frac{ (1-\rho)
\ln\left(1-\rho\right)}{\rho} - \frac{ (1-r_2)
\ln\left(1-r_2\right)}{r_2}\right] \ ,\label{nloa}
\end{eqnarray}
with
\begin{eqnarray}
&&\bar{u}=1-u,\ \ \rho = [r_1 +u(r_2 -r_1)-u(1-u)M_P^2/m_b^{2}],
\ \ \mbox{Li}_2(x)=-\int^x_0\frac{dt}t \ln(1-t), \nonumber \\
&&G\left(\rho \right) = \mbox{Li}_2(\rho) + \ln^2(1-\rho)
-\ln(1-\rho) \left(1 -\ln\frac{m_b^{2}}{\mu^2} \right) ,
\end{eqnarray}
where the dilogarithm function ${\rm Li}_2(x)=
-\int_0^x\frac{dt}{t}\ln(1-t)$, $r_1 = q^2/m_b^{2}$ and
$r_2=(p+q)^2/m_b^{2}$.

Next, the QCD LCSR for $f^{+}_{B\to P}(q^2)$ can be schematically
written as
\begin{equation}
f_{B} f^{+}_{B\to P}(q^2)=\frac{1}{2m_B^2}\int_{m_b^{2}}^{s_0}
e^{(m_B^2-s)/M^2}\left[\rho^{+}_{T2}(s,q^2)+\rho^{+}_{T4}(q^2)\right]
ds \;,
\end{equation}
where $\rho^{+}_{T2}(s,q^2)$ is the contribution from the twist-2 DA
and $\rho^{+}_{T4}(q^2)$ is for twist-4 DA, $f_{B}$ is the B-meson
decay constant. The Borel parameter $M^2$ and the continuum
threshold $s_0$ are determined such that the resulting form factor
does not depend too much on the precise values of these parameters;
in addition the continuum contribution, which is the part of the
dispersive integral from $s_0$ to $\infty$ that has been subtracted
from both sides of the equation, should not be too large, e.g. less
than $30\%$ of the total dispersive integral.

As for the LO twist-2 and twist-4 contributions, we obtain
\begin{eqnarray}
f_{B} f^{+}_{B\to P}(q^2)|_{LO} &=& \frac{m_b^{2} f_P}{m_B^2}
e^{\frac{m_B^2}{M^2}} \Bigg\{\int_{\triangle}^{1} du
e^{-\frac{m_b^{2}-\bar{u}(q^2-u m_P^2)}{u M^2}}
\left[\frac{\varphi_P(u)}
{u}+\frac{G_{4P}(u)}{u M^2}-\frac{m_b^{2} \phi_{4P}(u)}{4u^3 M^4}\right] \nonumber \\
&&+\int_{0}^{1} dv \int D\alpha_i\frac {\theta(\alpha_1+v \alpha_3
-\Delta)}{(\alpha_1+v \alpha_3)^2 M^2} e^{-\frac
{m_b^{2}-(1-\alpha_1-v\alpha_3)(q^2-(\alpha_1+v\alpha_3) m_P^2)}
{M^2 (\alpha_1+v\alpha_3)}}\times \nonumber\\
&& \left[2 \Psi_{4P}(\alpha_i) +2\tilde{\Psi}_{4P}(\alpha_i)
-\Phi_{4P}(\alpha_i)-\tilde{\Phi}_{4P}(\alpha_i)\right] \Bigg\},
\end{eqnarray}
where $\triangle=\frac{m_b^{2}-q^2}{s_0-q^2}$ for $M_P=0$;
$\triangle=\frac{\sqrt{(s_0-q^2-M_P^2)^2 +4M_P^2(m_b^{2}-q^2)}
-(s_0-q^2-M_P^2)}{2M_P^2}$ for $M_P\neq0$.

As for the NLO twist-2 contribution, it is convenient to write the
NLO $\rho^{+}_{T2}(s,q^2)$ in the following form
\begin{equation}
\rho^{+}_{T2}(s,q^2)|_{NLO}=-\frac{f_P}{\pi}\left(\frac{\alpha_s
C_F}{4\pi}\right)\int_0^1 du \phi_{P}(u,\mu) {\rm Im} T_1^+\; \left
(\frac{q^2}{m_b^{2}}, \frac{s}{m_b^{2}},u,\mu\right ) ,
\end{equation}
where
\begin{eqnarray}
\frac{1}{4 \pi}{\rm Im}_s T^{+}_1 &=& \theta(1-\rho ) \left [
\left.\frac{L_2(r_2)}{\rho -1}\right|_{+}+ \frac{
1-r_{1}}{(r_{2}-r_{1}) (r_{2}-\rho )} L_1(r_2)  -
\frac{r_{2}-1}{(r_2-\rho)r_2} \right ] \nonumber \\
&+& \theta (\rho -1)\left [\left.\frac{L_2(r_2) - 2 L_1(\rho)}{\rho
-1}\right |_{+} + \frac{ 1-r_{1}}{(r_{2}-r_{1}) (r_{2}-\rho )}
L_1(r_2)  \right . \nonumber \\
&+& \left.\frac{ 1 + \rho - r_{1}-r_{2}}{(r_{1}-\rho ) (r_{2}-\rho
)} L_1(\rho)+ \frac{1}{2\rho}\left ( 1 - \frac{1}{\rho} -
\frac{2}{r_2} \right )\right ] \nonumber\\
&+&\delta(\rho -1) \left [ \left (\ln\frac{r_2 -1}{1- r_1} \right
)^2 - \left (\frac{1-r_2}{r_2}+ \ln r_2 \right ) \ln
\frac{(r_2-1)^2}{1-r_1} \right . \nonumber \\
&-& \frac{3}{2} \ln \left(\frac{m_b^{2}}{\mu ^2}\right) \left . +
{\rm Li}_2(r_{1}) - 3{\rm Li}_2(1-r_{2})+3-\frac{\pi^2}{2}\right
]\,,
\end{eqnarray}
for the case of $r_1<1$ and $r_2>1$. The operation $``+"$ is defined
by
\begin{equation}
\left. \int d\rho f(\rho)\frac{1}{1-\rho}\right|_{+} =\int d\rho
[f(\rho)-f(1)]\frac{1}{1-\rho}.
\end{equation}
The two functions $L_1(x) = \ln \left
[\frac{(x-1)^2}{x}\frac{m_b^{2}}{\mu^2} \right ]- 1$ and $L_2(x) =
\ln \left [\frac{(x-1)^2}{x}\frac{m_b^{2}}{\mu^2} \right] -
\frac{1}{x}$ are introduced to make the formulae short. The above
formulae are derived in the Feynman gauge and by regularizing both
the ultraviolet and collinear divergences by the standard
dimensional regularization in the $\overline{MS}$ scheme.

\subsection{Calculation of $f^{0,T}_{B\to P}$ within LCSR}

For convenience, we calculate the combined function $f^{*}_{B\to
P}(q^2)=\left[f^{+}_{B\to P}(q^2)+f^{-}_{B\to P}(q^2)\right]$ first
and then derive $f^{0}_{B\to P}$ with the help of
Eq.(\ref{relation0p}). The OPE results for the needed invariant
amplitudes $\Pi^{-,T}$ can be represented as a sum of LO and NLO
parts:
\begin{eqnarray}
\Pi^{-,T}(q^2,(p+q)^2)&=& \Pi^{-,T}_0(q^2,(p+q)^2)+
\frac{\alpha_sC_F} {4\pi}\Pi^{-,T}_1(q^2,(p+q)^2),
\end{eqnarray}
where $\Pi^{-,T}_0(q^2,(p+q)^2)$ and $\Pi^{-,T}_1(q^2,(p+q)^2)$
stand for the LO and NLO contributions respectively. As for the LO
invariant amplitudes, we obtain:
\begin{eqnarray}
\Pi^{-}_{0}(q^2,(p+q)^2) &=& 2f_P m^2_b
\int^1_0\frac{du}{u^2}\frac{1}{\Delta^2}G_{4P}(u),\label{lop}\\
\Pi^{T}_{0}(q^2,(p+q)^2) &=& 2m_b f_P\left[\int_0^1 \frac{du}{u}
\frac{\varphi_P(u)}{\Delta} -\int_0^1 \frac{1}{4u^2\Delta^2}
\left(1+\frac{2m_b^2}{u\Delta}\right)\phi_{4P}(u)\right.\nonumber\\
&&\!\!\!\!\!\!\!\!\!\!\!\!\!\!\!\!\!\!\!\!\!\!\!\!\!\!\! \left.
+\int_0^1 dv \int {\cal D} \alpha_i
\frac{2\Psi_{4P}(\alpha_i)-(1-2v) \Phi_{4P}(\alpha_i) +
2(1-2v)\widetilde{\Psi}_{4P}(\alpha_i)-
\widetilde{\Phi}_{4P}(\alpha_i)}{\Delta^2(\alpha_1+v\alpha_3)^2}\right]
. \label{lot}
\end{eqnarray}
Similar to the case of $\Pi^{+}_{0}(q^2,(p+q)^2)$, one may also
observe that only the twist-2 and twist-4 contributions are
contained in the above expressions, and the twist-3 terms are
rightly eliminated by taking the present adopted chiral currents
within the correlators. The NLO invariant amplitude $\Pi^{-,T}_1$
for the twist-2 contribution can be written in the following
factorized form:
\begin{equation}
\Pi^{-,T}_1(q^2,(p+q)^2) = -f_P \int_0^1 du
T^{-,T}_1(q^2,(p+q)^2,u)\varphi_P(u),
\end{equation}
where by taking $m_b$ to be the b-quark pole mass, we have
\begin{eqnarray}
T^{-}_1(r_1,r_2,u)&=& \frac{2(r_1-r_2)[r_{1}+(1-r_1)\ln
(1-r_1)]}{r_1^2 (1-\rho)}+\frac{2(1-r_1) (r_1+r_2) \ln
(1-r_1)}{r_1^2 (r_1-\rho)}\nonumber \\
&+& \frac{4(1-r_2) \ln (1-r_2)}{r_2 (\rho-r_2)}-
\frac{2(1-\rho)(r_2+\rho ) \ln (1-\rho )}{u(\rho-r_2)\rho
^2}+\frac{2(r_1-r_2)}{r_1 \rho} \label{nlop}
\end{eqnarray}
and
\begin{eqnarray}
T^{T}_1(r_1,r_2,u)&=& \frac{4}{1-\rho}\left(3-2\ln\frac{m_b^{2}}
{\mu^2}\right) -\frac{4}{1-\rho}[2G(\rho)-G(r_1)-G(r_2)] \nonumber\\
&-&\frac{4}{(r_1-r_2)^2}\left(\frac{ 1-r_2}{u}[G(r_1)-G(\rho)]+
\frac{1-r_1}{\bar{u}}[G(r_2)-G(\rho)]\right)\nonumber\\
&-&\frac{4}{1-\rho}\left(\frac{ 1-r_2}{r_2}\ln (1-r_2)-
\frac{ 1-r_1}{r_1}\ln (1-r_1) \right) \nonumber\\
&+&4\left(\frac{1- r_1}{r_1 (\rho-r_1)}\right)
\ln\left(\frac{1-r_1}{1-\rho}\right) -4\left(\frac{1-r_2}{(\rho-r_2)
r_2}\right) \ln\left(\frac{1-r_2}{1-\rho}\right)\nonumber \\
&-&2\left(\frac{\rho+\ln (1-\rho )}{\rho ^2}\right) +\left(
\frac{-4r_1+ 2r_2 r_1+4r_2}{r_1 r_2 \rho }\right) \ln (1-\rho ).
\label{nlot}
\end{eqnarray}

Schematically, the QCD LCSRs for $f^{*,T}_{B\to P}$ can be written
as
\begin{eqnarray}
f_{B} f^{*}_{B\to P}(q^2) &=& \frac{1}{m_B^2}\int_{m_b^{2}}^{s_0}
e^{(m_B^2-s)/M^2}\left[\rho^{*}_{T2}(s,q^2)+\rho^{*}_{T4}(q^2)\right] ds \;,\\
f_{B} f^{T}_{B\to P}(q^2) &=& \frac{m_B +m_P}
{2m_B^2}\int_{m_b^{2}}^{s_0} e^{(m_B^2-s)/M^2}
\left[\rho^{T}_{T2}(s,q^2)+ \rho^{T}_{T4}(q^2) \right] ds \;.
\end{eqnarray}

As for the LO twist-2 and twist-4 contributions, with the help of
the Eqs.(\ref{lop},\ref{lot}), we obtain
\begin{eqnarray}
f_{B} f^{*}_{B\to P}(q^2)|_{LO} &=& \frac{2m_b^{2} f_P}{m_B^2}
e^{\frac{m_B^2}{M^2}} \int_{\triangle}^{1} du e^{-\frac{m_b^{2}
-\bar{u}(q^2-u m_P^2)}{u M^2}}\left[\frac{G_{4P}(u)}{u^2 M^2}\right] , \\
f_{B} f^{T}_{B\to P}(q^2)|_{LO} &=& \frac{(m_B +m_P)m_b^{} f_P}
{m_B^2} e^{\frac{m_B^2}{M^2}} \Bigg\{\int_{\triangle}^{1} du
e^{-\frac{m_b^{2}-\bar{u}(q^2-u m_P^2)}{u M^2}}\cdot\nonumber \\
&&\left[\frac{\varphi_P(u)} {u}- \frac{\phi_{4P}(u)}{4u^2
M^2}\left(1+\frac{m_b^{2}}{u
M^2}\right)\right]+\nonumber\\
&&\int_{0}^{1} dv \int D\alpha_i\frac {\theta(\alpha_1+v \alpha_3
-\Delta)}{(\alpha_1+v \alpha_3)^2 M^2} e^{-\frac
{m_b^{2}-(1-\alpha_1-v\alpha_3)(q^2-(\alpha_1+v\alpha_3) m_P^2)}
{M^2 (\alpha_1+v\alpha_3)}}\times \nonumber\\
&&\left[2\Psi_{4P}(\alpha_i)-(1-2v)\Phi_{4\pi}(\alpha_i) +
2(1-2v)\widetilde{\Psi}_{4P}(\alpha_i)-
\widetilde{\Phi}_{4P}(\alpha_i)\right] \Bigg\} .
\end{eqnarray}
From the above equations, we immediately obtain the relations among
$f^{\pm,T}_{B\to P}(q^2)$ at the LO and up to the twist-3 accuracy,
i.e.
\begin{equation}\label{relationat}
f^{-}_{B\to P}(q^2)=-f^{+}_{B\to P}(q^2) \;\; {\rm and} \;\;
f^{T}_{B\to P}(q^2)=\frac{m_B + m_P} {m_b} f^{+}_{B\to P}(q^2) \;,
\end{equation}
which agree with the conclusions drawn in Ref.\cite{huangz}.
Moreover, with the help of Eqs.(\ref{relation0p},\ref{relationat}),
we obtain
\begin{equation}\label{relation0a}
f^{0}_{B\to P}(q^2)=\left[1-\frac{q^2}{m_B^2
-m_P^2}\right]f^{+}_{B\to P}(q^2) .
\end{equation}

As for the NLO twist-2 contribution, the NLO
$\rho^{*,T}_{T2}(s,q^2)$ can be written as
\begin{equation}
\rho^{*,T}_{T2}(s,q^2)|_{NLO}=-\frac{f_P}{\pi}\left(\frac{\alpha_s
C_F}{4\pi}\right)\int_0^1 du \phi_{P}(u,\mu) {\rm Im} \;
T^{-,T}_{1}\left (\frac{q^2}{m_b^{2}}, \frac{s}{m_b^{2}},u,\mu\right
) ,
\end{equation}
where
\begin{eqnarray}
\frac{1}{2\pi}{\rm Im}_s T^{-}_{1} &=&\theta (1-\rho ) \left[
\frac{2 (1-r_{2}) }{r_{2} (r_{2}-\rho )} \right ]- \frac{\theta
(\rho-1)}{r_1 - \rho} \left [ \frac{r_1 - r_2}{\rho^2} - \frac{(2 -
r_2)(r_2 - r_1)}{r_2 \rho}\right.  \nonumber \\
&+& \left.\frac{2 (r_2-1)}{r_2} \right ]+\delta (\rho -1)
\left[1-\frac{r_{2}}{r_{1}}- \frac{(r_{1}-1) (r_{1}-r_{2}) \ln
(1-r_{1})}{r_{1}^2}\right ]\,,
\end{eqnarray}
and
\begin{eqnarray}
\frac{1}{4\pi}{\rm Im}_s T_1^{T} &=&\theta (1-\rho ) \left [\left.
\frac{L_2(r_2)}{\rho-1} \right|_{+} - \frac{1-r_1}
{(r_2-r_1)(\rho -r_2)}L_1(r_2)- \frac{r_2-1}{r_2 (\rho - r_2) }\right ]\nonumber \\
&+& \theta(\rho-1) \left [\left.\frac{L_2(r_2)-2L_1(\rho)}{\rho-1}
\right |_{+} - \frac{1-r_1}{(r_2-r_1)(\rho -r_2)}L_1(r_2) \right. \nonumber \\
&-& \left. \frac{r_1+r_2 -\rho -1}{(r_1-\rho)(r_2-\rho)}L_1(\rho)-
\frac{r_1-1}{r_1(\rho-r_1)} - \frac{2(r_2-r_1) + r_1
r_2}{2 r_1 r_2 \rho} + \frac{1}{2 \rho^2}\right ]\nonumber \\
&+& \delta (\rho -1) \left[ \left ( \ln  \frac{r_{2}-1}{1 - r_1}
\right )^2 - \ln\frac{(r_{2}-1)^2}{1-r_{1}} \left(\ln
r_{2}+\frac{1}{r_{2}}-1\right) +3 -\frac{\pi^2}{2} \right . \nonumber \\
&-& \left . \left(1-\frac{1}{r_{1}}\right) \ln (1-r_{1}) -2 \ln
\left(\frac{m_b^{2}}{\mu ^2}\right)+{\rm Li}_2(r_{1})- 3{\rm
Li}_2(1-r_{2})\right]\,,
\end{eqnarray}
for the case of $r_1<1$ and $r_2>1$.

As a cross check of the above NLO formulae for the twist-2
contributions, it can be found that our present results for
$f^{+,*,T}_{B\to\pi}$ agree with Ref.\cite{duplan} by transforming
the formulae for the $\overline{MS}$ $b$-quark mass to be the ones
for the $b$-quark one-loop pole mass, except for an overall factor
{\it 2} \footnote{The overall factor {\it 2} comes from the
different choices of correlation function.}.

Here similar to the treatment of Refs.\cite{Kho3,huang3,wk1}, we
have adopted the $b$-quark pole mass to do the calculation.
Refs.\cite{duplan,duplan2} have argued to use the $b$-quark
$\overline{MS}$ running mass other than the pole mass. Numerically,
we shall show in due places that if properly choosing the possible
ranges for the undetermined parameters, these two treatments are in
fact equivalent to each other within reasonable uncertainties. We
prefer to take the pole quark mass, sine the pole quark mass is
universal that can be determined through proper potential model
analysis or through lattice QCD calculation, while the running quark
mass is process dependent, i.e. depends on the renormalization
scheme and the renormalization scale of a particular process.

\section{Numerical results for $f^{+,0,T}_{B\to\pi,\; K}(q^2)$
within the QCD LCSR with chiral current}

\subsection{Parameters and distribution amplitudes of the light
mesons}

First, we specify the input parameters used in the LCSRs for
$B\to\pi$ and $B\to K$ transition form factors. For the needed meson
masses and the light mesons' decay constants, we adopt the center
values as listed by the Particle Data Group \cite{pdg}
\begin{eqnarray}
&& f_\pi=130.4 {\rm MeV},\; f_K=155.5{\rm MeV},\nonumber\\
&& M_B=5.279{\rm GeV} ,\; M_\pi=139.570{\rm MeV},\; M_K=493.667{\rm
MeV}. \nonumber
\end{eqnarray}

\begin{table}
\centering
\begin{tabular}{|c||c|c|c|}
\hline ~~~ - ~~~ & ~~~$s_0$ (GeV$^2$)~~~ & ~~~$M^2$~~~ & ~~~ $f_B$ (GeV) ~~~\\
\hline\hline
$m_b=4.75$ (GeV) & 33.0 & 2.48 & 0.192 \\
\hline
$m_b=4.80$ (GeV) & 32.6 & 2.28 & 0.169 \\
\hline
$m_b=4.85$ (GeV) & 32.3 & 2.10 & 0.146 \\
\hline
\end{tabular}
\caption{The value of $f_B$ (in units: GeV) within the LCSRs with
chiral currents up to NLO, the corresponding formulae can be found
in Ref.\cite{wk1}, where $m_b$ is taken to be the $b$-quark pole
mass.} \label{tabfb}
\end{table}

As has been argued in the last section, we shall adopt the $b$-quark
pole mass to do numerical calculation throughout the paper. As for
the value of $f_B$, to be consistent with the present calculation
technology, they should be determined by using the two-point sum
rule with proper chiral currents up to NLO. Such a calculation has
been done in Ref.\cite{wk1}, the interesting reader may turn to
Ref.\cite{wk1} for more calculation detail, and here we only quote
some typical results as shown in Tab.\ref{tabfb}, where the one-loop
pole mass $m_b$ is taken to be $(4.80\pm0.05)$ GeV \cite{mbmass}.

\begin{figure}
\includegraphics[width=0.5\textwidth]{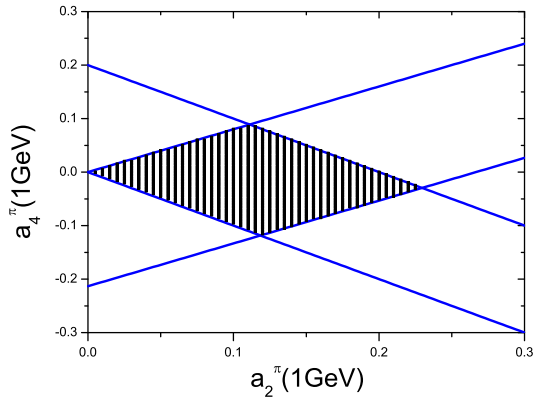}\\
\caption{$a_2^\pi(1 \mbox{GeV})$ and $a_4^{\pi}(1 \mbox{GeV})$ as
determined from the two constraints adopted in the body of the text,
where the rhomboid stands for the allowable range.} \label{a24pi}
\end{figure}

Naively, the leading twist-2 DAs $\phi_\pi$ and $\phi_K$ can be
expanded as Gegenbauer polynomials as shown in APPENDIX A. The first
two Gegenbauer moments, e.g. $a_2^\pi$ and $a_4^\pi$ for pion and
$a^K_1$ and $a^K_2$ for kaon, have been studied with various
processes. We adopt two constraints for $a_2^\pi(1 \mbox{GeV})$ and
$a_4^{\pi}(1 \mbox{GeV})$, e.g. $a_2^{\pi}(1 \mbox{GeV})
+a_4^{\pi}(1 \mbox{GeV})=0.1\pm 0.1$ \cite{Bakulev} and
$-\frac{9}{4}a_2^{\pi}(1 \mbox{GeV}) +\frac{45}{16}a_4^{\pi}(1
\mbox{GeV})+\frac{3}{2}=1.2\pm 0.3$ \cite{pball2,braun0}, such that
the allowed values of $a^\pi_2$ and $a^\pi_4$ are correlated and
given by the rhomboid shown in Fig.(\ref{a24pi}). Note here we do
not adopt the wider range of $a^\pi_2=0.25\pm0.15$ as suggested by
Ref.\cite{ballmoments}, since we prefer a more asymptotic-like pion
DA as favored by a very recent QCD LCSR analysis of $B\to\pi$ vector
form factor \cite{wu}. The first Gegenbauer moment $a_1^K$ has been
studied by several references, e.g.
Refs.\cite{quark1,kaonlcsr,pballa1k,ballmoments,lattice1,lattice2}
and etc. For convenience, we quote the values for the twist-2
Gegenbauer moments of kaon as obtained from the average of those
obtained in literature to do the discussion, $a_1^K(1
\mbox{GeV})=0.06\pm0.03$ and $a_2^K(1 \mbox{GeV})=0.25\pm0.15$
\cite{ballmoments}.

\begin{table}
\begin{tabular}{|c||c|c|c|c|c|}
\hline
$a^\pi_2(\mu_0)$  & \multicolumn{1}{|c|}{$0.00$} & \multicolumn{3}{|c|}{$0.115$} & \multicolumn{1}{|c|}{$0.230$}\\
\hline
$a^\pi_4(\mu_0)$   & ~~~$0.00$~~~ & ~~~ $0.092$~~~ &~~~ $-0.015$~~~ &~~~ $-0.120$ ~~~& ~~~ $-0.030$ ~~~\\
\hline\hline
$A_{\pi}(GeV^{-2})$ & 226.0 & 196.7 & 199.4  & 199.1  & 173.8  \\
\hline
$B_{\pi}$  &  -0.079 & -0.024 & -0.018  &  -0.014& 0.043 \\
\hline
$C_{\pi}$ & 0.027 & 0.073 & 0.012  & -0.050 & -0.00656  \\
\hline
$\beta_{\pi} (GeV)$  & 0.902 &  0.862 & 0.868  & 0.870 &  0.832 \\
\hline
\end{tabular}
\caption{Pion twist-2 wavefunction parameters for some typical
Gegenbauer moments, where $\mu_0=1GeV$. Note the obtained WF
parameters are for $\mu=2.2GeV$. \label{tabpionpara}}
\end{table}

\begin{table}
\begin{tabular}{|c||c|c|c|c|c|c|c|c|c|}
\hline
$a^K_1(\mu_0)$  & \multicolumn{3}{|c|}{$0.09$} & \multicolumn{3}{|c|}{$0.06$} & \multicolumn{3}{|c|}{$0.03$}\\
\hline
$a^K_2(\mu_0)$   & $0.40$ &  $0.25$ & $0.10$ & $0.40$ &  $0.25$ & $0.10$ & $0.40$ &  $0.25$ & $0.10$\\
\hline\hline
$A_{K}(GeV^{-2})$ & 171.2 & 209.3 &253.8& 172.6 & 211.8 & 255.9 & 173.9&  213.5& 258.1 \\
\hline
$B_{K}$  & 0.0845  & 0.0732 & 0.0588& 0.107 & 0.0966 & 0.0825 & 0.130& 0.119& 0.106 \\
\hline
$C_{K}$ & 0.203 & 0.122 & 0.0371& 0.207 & 0.127 & 0.0422 & 0.211 & 0.132& 0.0471 \\
\hline
$\beta_{K} (GeV)$  & 0.774 & 0.821 & 0.869&  0.775 & 0.823 & 0.870 & 0.776& 0.824 & 0.871 \\
\hline
\end{tabular}
\caption{Kaon twist-2 wavefunction parameters for some typical
Gegenbauer moments, where $\mu_0=1GeV$. Note the obtained WF
parameters are for $\mu=2.2GeV$. \label{tabkaonpara}}
\end{table}

\begin{figure}
\includegraphics[width=0.60\textwidth]{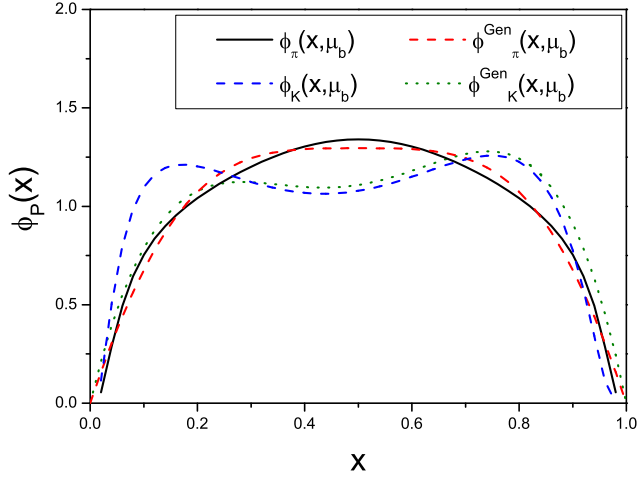}\\
\caption{Typical distribution amplitudes $\phi_P(x)$ at $\mu_b=2.2
\mbox{GeV}$, where $\phi_\pi(x)$ and $\phi_K(x)$ are for $a^\pi_2(1
\mbox{GeV})=0.115$ and $a^\pi_4(1 \mbox{GeV})=-0.015$, $a_1^K(1
\mbox{GeV})=0.06$ and $a_2^K(1 \mbox{GeV})=0.25$ respectively, and
$\phi^{gen}_\pi(x)$ and $\phi^{gen}_K(x)$ are for Gegenbauer
expansion (\ref{phigen}) with the same Gegenbauer moments.}
\label{phip}
\end{figure}

\begin{table}
\begin{tabular}{|c|c|c||c|c|}
\hline
twist & $\pi$ & $\mu=1 $ GeV & $K$ & $\mu=1 $ GeV\\
\hline \hline
4 & $\delta^2_{\pi}$& $0.18\pm 0.06$  GeV$^2$ & $\delta^2_{K}$& $0.20\pm 0.06$  GeV$^2$\\
  & $\epsilon_{\pi}$&$ \frac{21}{8}(0.2\pm 0.1)$& $\epsilon_{K}$&$ \frac{21}{8}(0.2\pm 0.1)$\\
\hline
\end{tabular}
\caption{Input parameters for the pion and kaon twist-4 DAs'
\cite{ballmoments}. \label{tabDA}}
\end{table}

Furthermore, for the twist-2 DAs, we do not adopt the Gegenbauer
expansion (\ref{phigen}), since its higher Gegenbauer moments are
still determined with large errors whose contributions may not be
too small, i.e. their contributions are comparable to that of higher
twist structures \cite{wk1}. As a compensation, we adopt the
suggestion of deriving the pion and kaon DAs from their
corresponding WFs by integrating over the transverse momentum
\cite{wk1}. And the twist-2 pion and kaon WFs can be constructed on
their first two Gegenbauer moments and on the BHL prescription
\cite{bhl}, i.e.
\begin{equation}\label{pimodel}
\Psi_{\pi}(x,\mathbf{k}_\perp) = [1+B_\pi C^{3/2}_2(2x-1)+C_\pi
C^{3/2}_4(2x-1)]\frac{A_\pi}{x(1-x)} \exp \left[-\beta_\pi^2
\left(\frac{\mathbf{k}_\perp^2+m_q^2}{x(1-x)}\right)\right],
\end{equation}
and
\begin{equation}\label{kmodel}
\Psi_{K}(x,\mathbf{k}_\perp) = [1+B_K C^{3/2}_1(2x-1)+C_K
C^{3/2}_2(2x-1)]\frac{A_K}{x(1-x)} \exp \left[-\beta_K^2
\left(\frac{\mathbf{k}_\perp^2+m_q^2}{x}+
\frac{\mathbf{k}_\perp^2+m_s^2} {1-x}\right)\right],
\end{equation}
where $q=u,\; d$, $C^{3/2}_{1,2}(1-2x)$ are Gegenbauer polynomials.
The constitute quark masses are set to be: $m_q=0.30{\rm GeV}$ and
$m_s=0.45{\rm GeV}$. After doing the integration over the transverse
momentum dependence, we obtain the twist-2 kaon DA, e.g. $
\phi_K(x,\mu_b)=\int_{k_\perp^2<\mu_b^2} \frac{d^{2}{\bf
k}_{\perp}}{16\pi^3} \Psi_K(x,{\bf k}_{\perp})$, where $\mu_b=2.2$
GeV for the present case. The Gegenbauer moments
$a^{\pi,K}_n(\mu_b)$ is defined as
\begin{equation}\label{moments}
a^{\pi,K}_n(\mu_b)=\frac{\int_0^1 dx
\phi_{\pi,K}(1-x,\mu_b)C^{3/2}_n(2x-1)} {\int_0^1 dx 6x(1-x)
[C^{3/2}_n(2x-1)]^2} \;.
\end{equation}
The four unknown parameters can be determined by the first two
Gegenbauer moments, the normalization condition $\int^1_0 dx
\int_{k_\perp^2<\mu_b^2} \frac{d^{2}{\bf
k}_{\perp}}{16\pi^3}\Psi_{\pi,K}(x,{\bf k}_{\perp}) =1$, and the
constraint $\langle \mathbf{k}_\perp^2 \rangle^{1/2}_K \approx
\langle \mathbf{k}_\perp^2 \rangle^{1/2}_\pi=0.350{\rm GeV}$
\cite{gh}, where the average value of the transverse momentum
square is defined as
\begin{displaymath}
\langle \mathbf{k}_\perp^2 \rangle^{1/2}_{\pi,K}=\frac{\int dx
d^2\mathbf{k}_\perp |\mathbf{k}_\perp^2| |\Psi_{\pi,K}(x,{\bf
k}_{\perp})|^2} {\int dx d^2\mathbf{k}_\perp |\Psi_{\pi,K}(x,{\bf
k}_{\perp})|^2} .
\end{displaymath}
Some typical parameters for the pion and kaon WFs are presented in
Tab.\ref{tabpionpara} and Tab.\ref{tabkaonpara}. A comparison with
the conventional Gegenbauler expansion DAs is presented in
Fig.(\ref{phip}). The remaining parameters for the twist-4 DA's
($\delta_{\pi,K}^2$, $\epsilon_{\pi,K}$) are presented in
Tab.\ref{tabDA}, which are taken from \cite{ballmoments}.

\subsection{Properties of $f^{+,0,T}_{B\to\pi,\; K}(q^2)$ within QCD LCSR with chiral
current}

Taking the above mentioned parameters, we discuss the properties of
$f^{+,0,T}_{B\to\pi,\; K}(q^2)$ within QCD LCSRs with chiral
current. At the maximum recoil region, $q^2=0$, by varying the
parameters within their reasonable regions, we obtain
\begin{eqnarray}\label{ffbpi}
f^{+,0}_{B\to\pi}(0)=0.260^{+0.059}_{-0.040}\;,\;\;f^{T}_{B\to\pi}(0)=0.276^{+0.052}_{-0.039}
\end{eqnarray}
and
\begin{eqnarray}\label{ffbk}
f^{+,0}_{B\to K}(0)=0.334^{+0.094}_{-0.069} \;,\;\;f^{T}_{B\to K}(0)
=0.379^{+0.092}_{-0.077} .
\end{eqnarray}

By comparing $B\to K$ form factors with the $B\to\pi$ form factors,
we find the following $SU_f(3)$-breaking effects among the $B\to$
light form factors:
\begin{equation}
\frac{f^{+,0}_{B\to K}(0)}{f^{+,0}_{B\to\pi}(0)}=
1.28^{+0.06}_{-0.08} \;,\;\; \frac{f^{T}_{B\to
K}(0)}{f^{T}_{B\to\pi}(0)}=1.37^{+0.07}_{-0.02} .
\end{equation}
It is found that this larger $SU_f(3)$-breaking effect is obtained
by taking a larger $a^K_2(1GeV)\in [0.10, 0.40]$; if taking a
smaller $a^K_2(1GeV)$, then one can obtain a smaller
$SU_f(3)$-breaking effect, e.g. $\frac{f^{+,0}_{B\to
K}(0)}{f^{+,0}_{B\to\pi}(0)}= 1.13\pm0.03$ for $a^K_2(1GeV) \in
[0.05,0.10]$ \cite{wk1} and $\frac{f^{+,0}_{B\to
K}(0)}{f^{+,0}_{B\to\pi}(0)}= 1.08^{+0.19}_{-0.17}$ for $a^K_2(1GeV)
\in [-0.11,0.27]$ \cite{kaonlcsr2}. Note that this larger
$SU_f(3)$-breaking effect is consistent with the some other LCSR
calculation as Refs.\cite{duplan2,kaonlcsr} and a recently
relativistic treatment that is based on the study of the
Dyson-Schwinger equation in QCD, i.e. $\frac{f^{+,0}_{B\to
K}(0)}{f^{+,0}_{B\to\pi}(0)}= 1.23$ \cite{dyson}. So a better
determination of $a^K_2(1GeV)$ will be helpful to obtain a better
understanding of the $SU_f(3)$-breaking effect.

\begin{figure}
\includegraphics[width=0.3\textwidth]{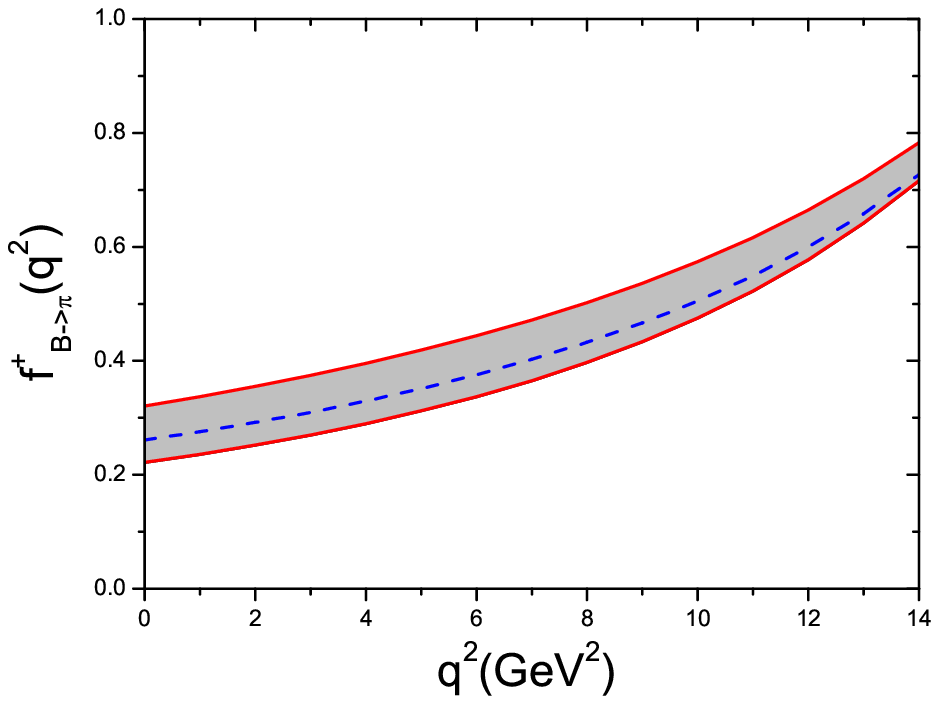}
\includegraphics[width=0.3\textwidth]{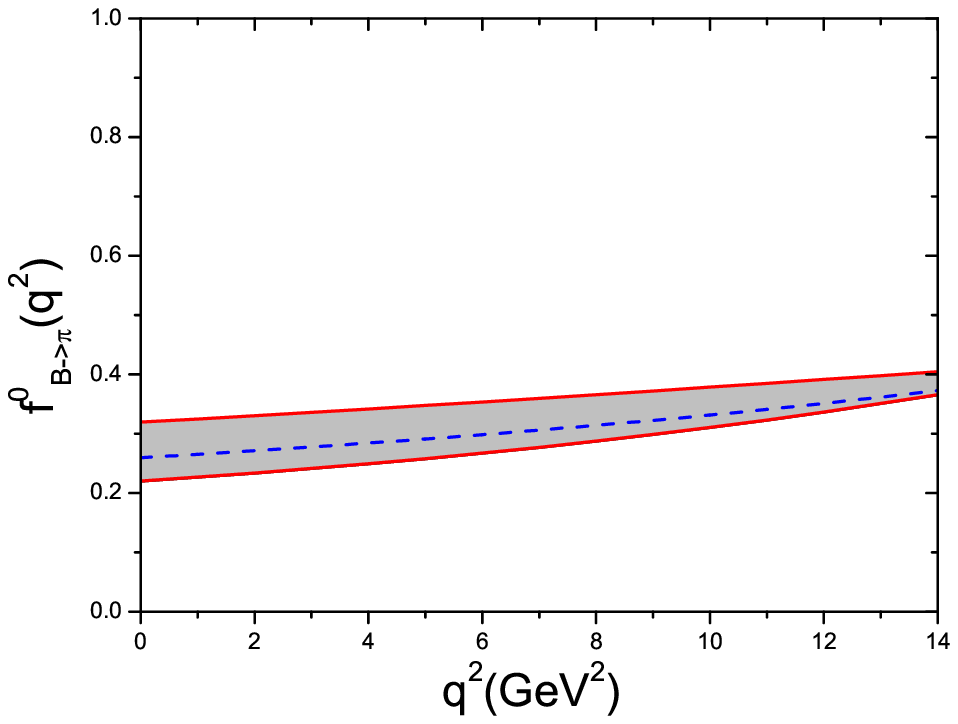}
\includegraphics[width=0.3\textwidth]{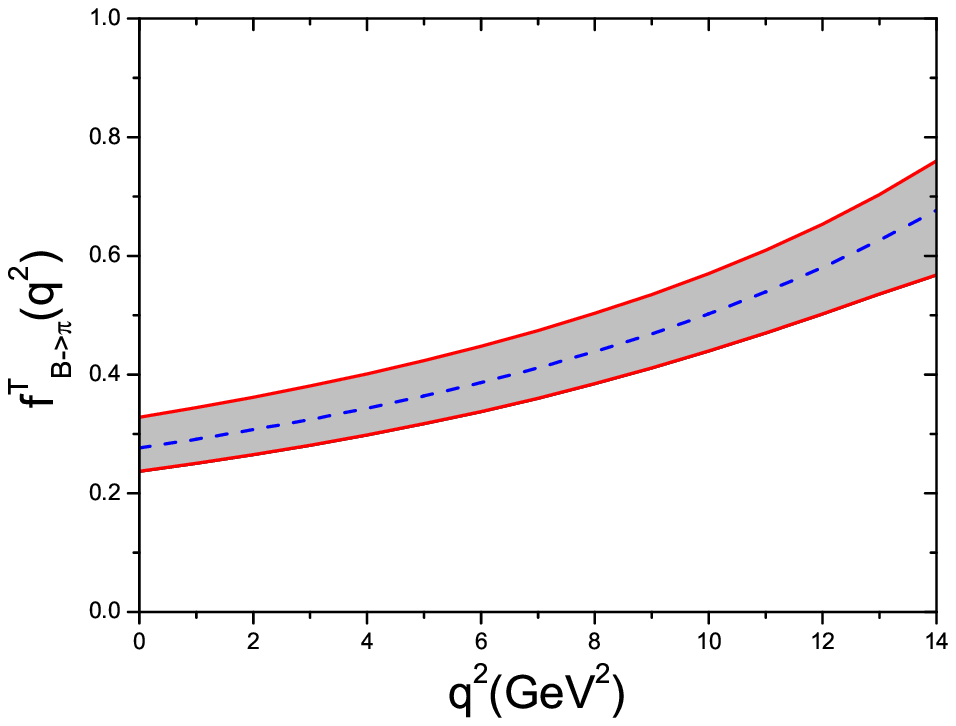}
\caption{Uncertainties of the $B\to\pi$ form factors
$f^{+,0,T}_{B\to\pi}(q^2)$ within the allowable regions for the
undetermined parameters. The center dashed line is for $m_b=4.80$
GeV, $a^\pi_2(1GeV)=0.115$, $a^\pi_4(1GeV)=-0.015$,
$\delta_\pi^2=0.18$ GeV$^2$ and $\epsilon_\pi=0.525$.} \label{fpi}
\end{figure}

\begin{figure}
\includegraphics[width=0.3\textwidth]{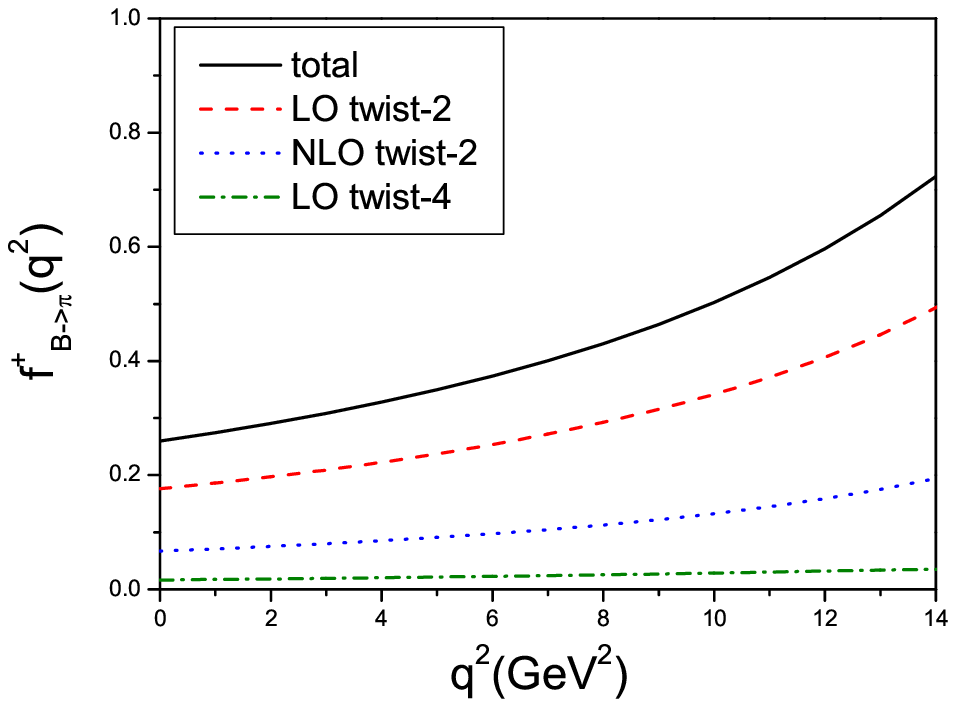}
\includegraphics[width=0.3\textwidth]{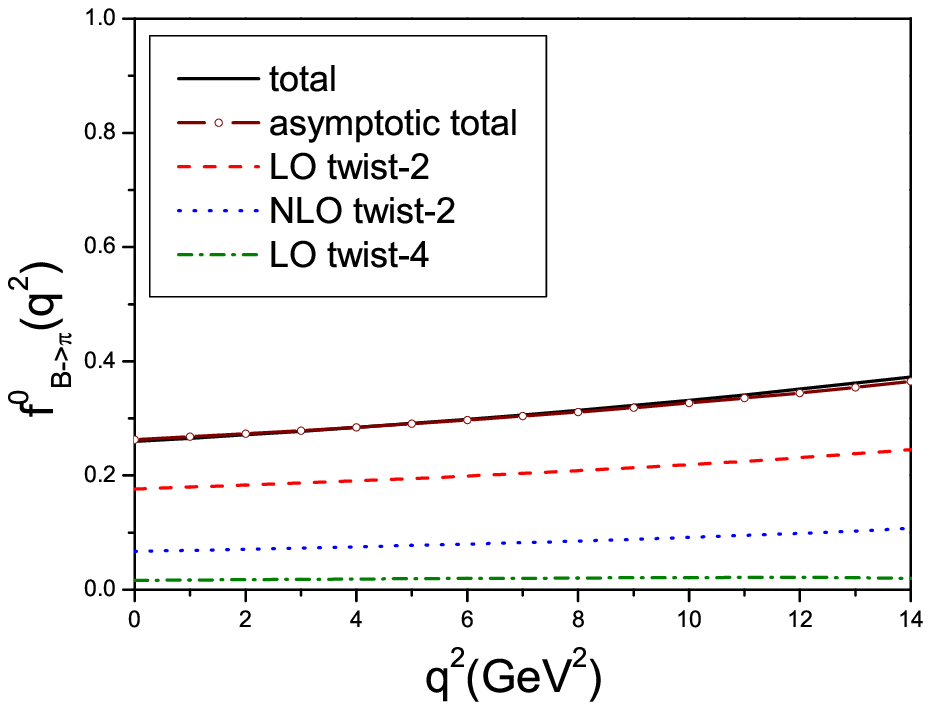}
\includegraphics[width=0.3\textwidth]{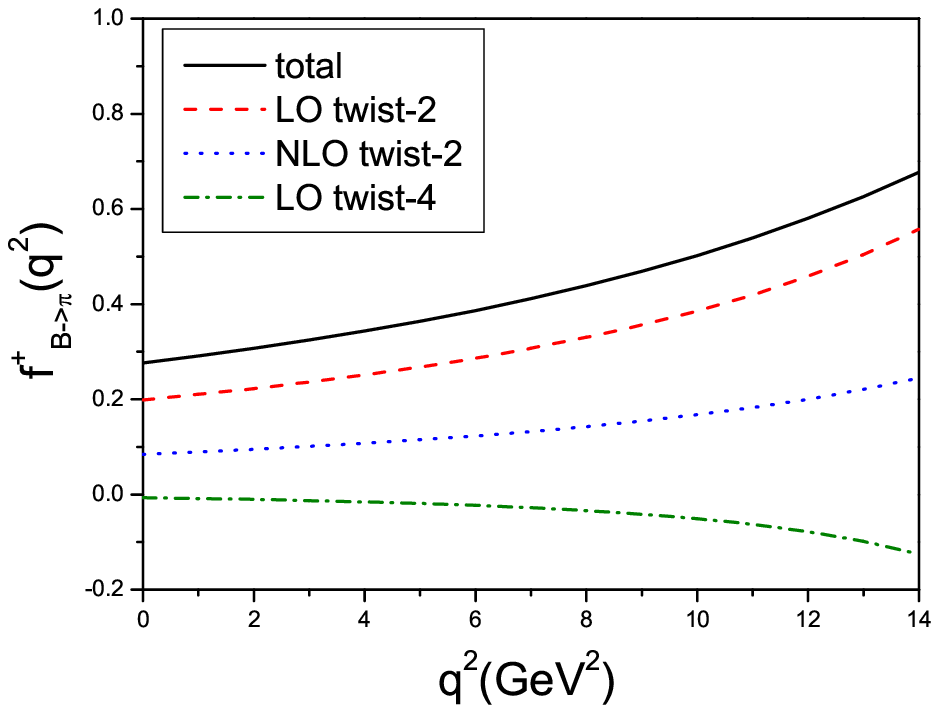}
\caption{Different parts' contributions to the $B\to\pi$ form
factors $f^{+,0,T}_{B\to\pi}(q^2)$ for all the parameters taken to
be their center values. The curve of asymptotic total in the middle
figure stands for the LO $f^{0}_{B\to\pi}(q^2)$ up to twist-3 that
is derived from Eq.(\ref{relation0a}).} \label{fpisep}
\end{figure}

We show the $B\to\pi$ vector, scalar and tensor form factors with
their corresponding errors in Fig.(\ref{fpi}), where the center
dashed line is for $m_b=4.80$ GeV, $a^\pi_2(1GeV)=0.115$,
$a^\pi_4(1GeV)=-0.015$, $\delta_\pi^2=0.18$ GeV$^2$ and
$\epsilon_\pi=0.525$. For $f^{+,0}_{B\to\pi}(q^2)$, the lower edge
of the shaded band is obtained by setting $m_b=4.75$ GeV,
$a^\pi_2(1GeV)=0.0$, $a^\pi_4(1GeV)=0.0$, $\delta_\pi^2=0.12$
GeV$^2$ and $\epsilon_\pi=0.2625$ and the upper edge is obtained by
setting $m_b=4.85$ GeV, $a^\pi_2(1GeV)=0.230$,
$a^\pi_4(1GeV)=-0.030$, $\delta_\pi^2=0.24$ GeV$^2$ and
$\epsilon_\pi=0.7875$. While for $f^{T}_{B\to\pi}(q^2)$, the lower
edge of the shaded band is obtained by setting $m_b=4.75$ GeV,
$a^\pi_2(1GeV)=0.0$, $a^\pi_4(1GeV)=0$, $\delta_\pi^2=0.24$ GeV$^2$
and $\epsilon_\pi=0.7875$ and the upper edge is obtained by setting
$m_b=4.85$ GeV, $a^\pi_2(1GeV)=0.230$, $a^\pi_4(1GeV)=-0.030$,
$\delta_\pi^2=0.12$ GeV$^2$ and $\epsilon_\pi=0.2625$. This
difference is caused by the fact that the twist-4 structures lead to
positive and negative contributions to the $f^{+,0}_{B\to\pi}(q^2)$
and $f^{T}_{B\to\pi}(q^2)$ respectively. The main uncertainties of
the form factors are caused by the value of $m_b$ and $a^\pi_2$, and
it can be found that all the $B\to\pi$ form factors shall increase
with the increment of $m_b$ and $a^\pi_2$. Further more, we obtain
$f^{T}_{B\to\pi}(0)/f^{+}_{B\to\pi}(0) \in [1.03,1.08]$. This shows
that the NLO correction shall affect the usual simple relation
(\ref{relationat}), e.g. $[f^{T}_{B\to\pi}(0)/f^{+}_{B\to\pi}(0)]=
(m_B +m_\pi)/m_b \in [1.12,1.14]$, to a certain degree. Furthermore,
we show contributions to the $B\to\pi$ form factors
$f^{+,0,T}_{B\to\pi}(q^2)$ from the different parts in
Fig.(\ref{fpisep}), where all the parameters are taken to be their
center values. For $f^{+,0}_{B\to\pi}(q^2)$, it can be found that
the LO twist-2, the NLO twist-2 and the LO twist-4 contributions are
positive, more specifically at $q^2=0$, they are about $68\%$,
$26\%$ and $6\%$ respectively. $f^{0}_{B\to\pi}(q^2)$ is very close
to the asymptotic LO result derived from Eq.(\ref{relation0a}),
which is caused by the fact that the LO twist-2 gives zero
contribution to the sum of the form factor
$[f^{+}_{B\to\pi}+f^{-}_{B\to\pi}]$ and then
$[f^{+}_{B\to\pi}+f^{-}_{B\to\pi}]$ gives negligible contribution to
$f^{0}_{B\to\pi}(q^2)$. For $f^{T}_{B\to\pi}(q^2)$, the LO twist-2,
the NLO twist-2 give positive contribution while the LO twist-4
gives negative contribution, more specifically at $q^2=0$, they are
about $72\%$, $30\%$ and $-2\%$ respectively.

\begin{figure}
\includegraphics[width=0.3\textwidth]{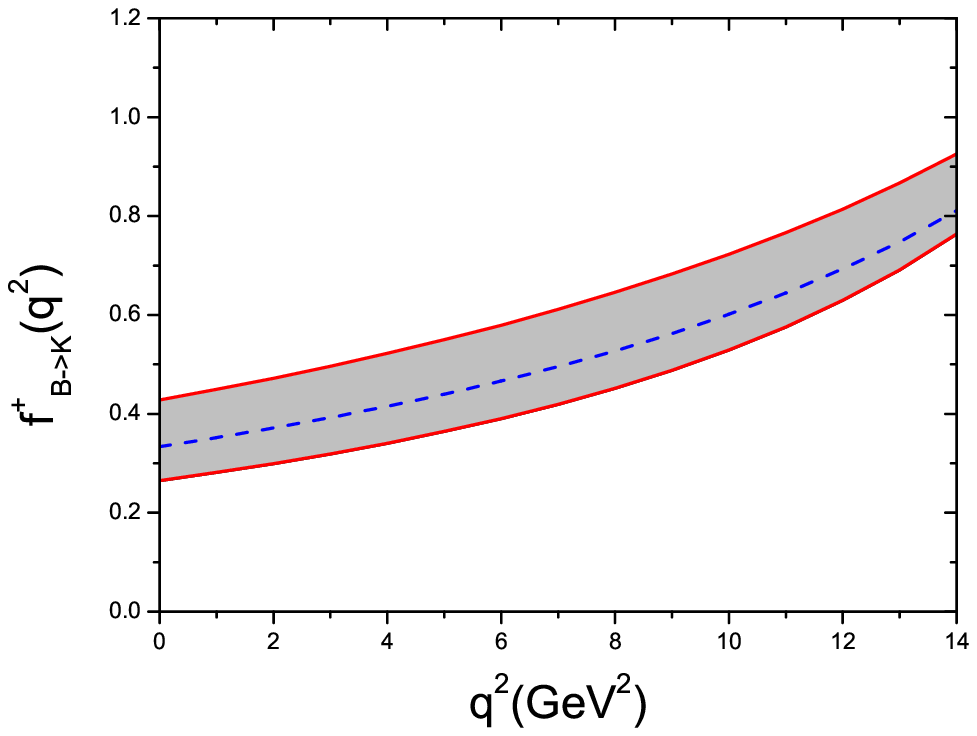}
\includegraphics[width=0.3\textwidth]{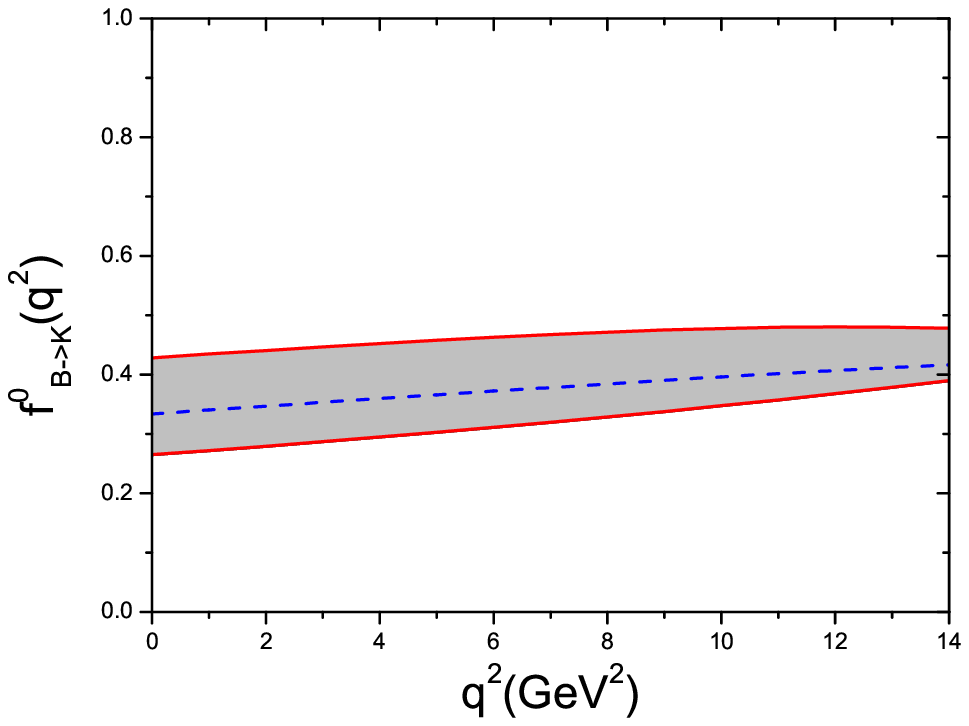}
\includegraphics[width=0.3\textwidth]{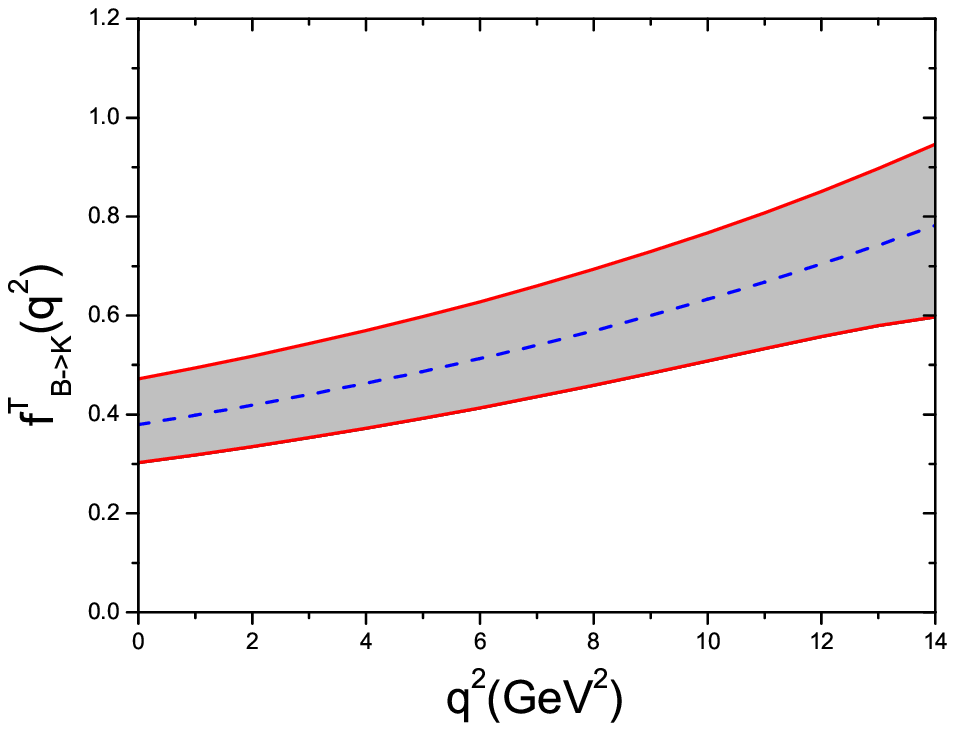}
\caption{Uncertainties of the $B\to K$ form factors
$f^{+,0,T}_{B\to\pi}(q^2)$ within the allowable regions for the
undetermined parameters. The center dashed line is for $m_b=4.80$
GeV, $a^K_1(1GeV)=0.06$, $a^K_2(1GeV)=0.25$, $\delta_K^2=0.20$
GeV$^2$ and $\epsilon_K=0.525$.} \label{fk}
\end{figure}

\begin{figure}
\includegraphics[width=0.3\textwidth]{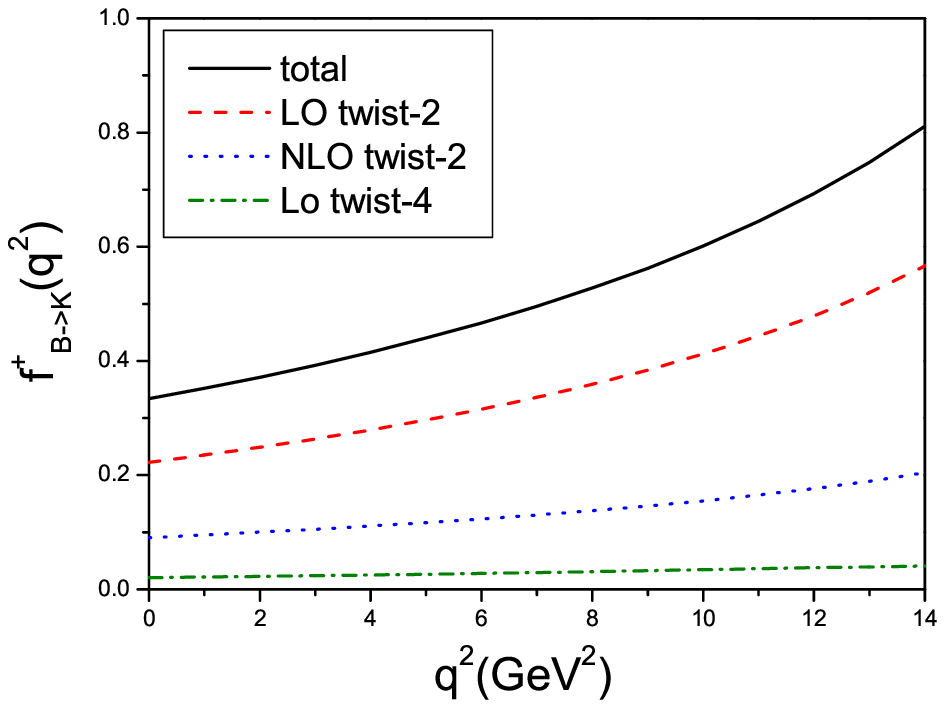}
\includegraphics[width=0.3\textwidth]{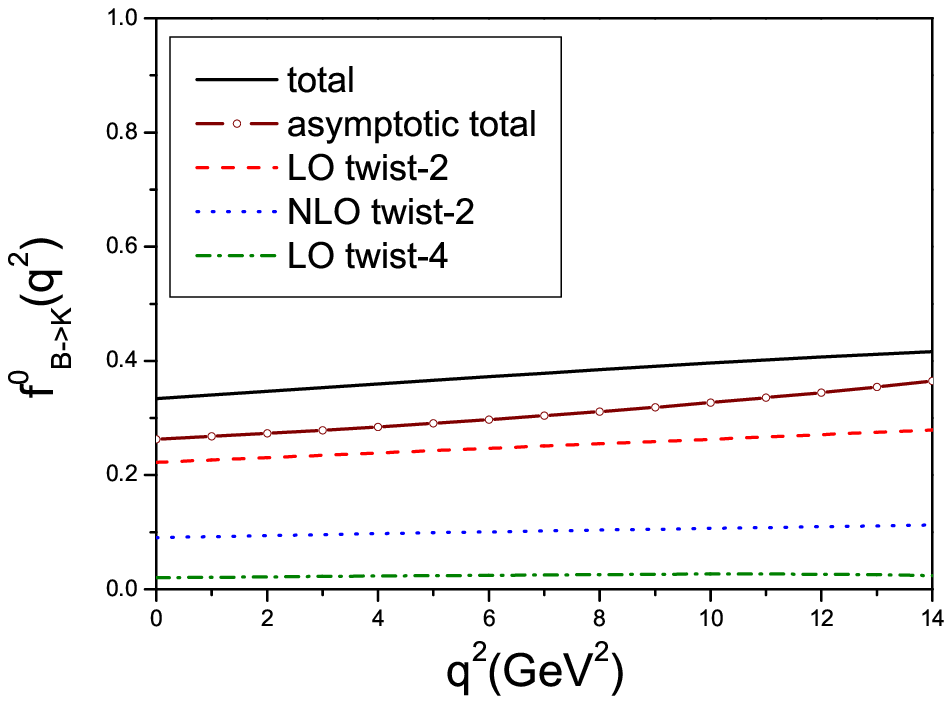}
\includegraphics[width=0.3\textwidth]{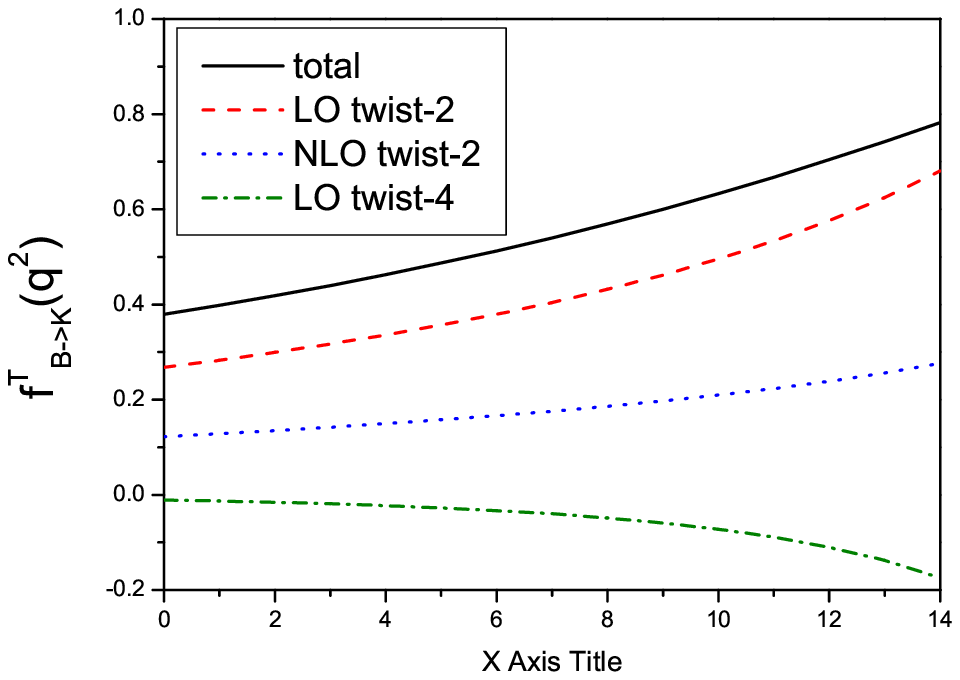}
\caption{Different parts' contributions to the $B\to K$ form factors
$f^{+,0,T}_{B\to K}(q^2)$ for all the parameters taken to be their
center values. The curve of asymptotic total in the middle figure
stands for the LO $f^{0}_{B\to K}(q^2)$ up to twist-3 that is
derived from Eq.(\ref{relation0a}).} \label{fksep}
\end{figure}

Second, we show the $B\to K$ form factors with their corresponding
errors in Fig.(\ref{fk}), where the center dashed line is for
$m_b=4.80$ GeV, $a^K_1(1GeV)=0.06$, $a^K_2(1GeV)=0.25$,
$\delta_K^2=0.20$ GeV$^2$ and $\epsilon_K=0.525$. For $f^{+,0}_{B\to
K}(q^2)$, the lower edge of the shaded band is obtained by setting
$m_b=4.75$ GeV, $a^K_1(1GeV)=0.09$, $a^K_2(1GeV)=0.10$,
$\delta_K^2=0.14$ GeV$^2$ and $\epsilon_K=0.2625$ and the upper edge
is obtained by setting $m_b=4.85$ GeV, $a^K_1(1GeV)=0.03$,
$a^K_2(1GeV)=0.40$, $\delta_K^2=0.26$ GeV$^2$ and
$\epsilon_K=0.7875$. While for $f^{T}_{B\to K}(q^2)$, the lower edge
of the shaded band is obtained by setting $m_b=4.75$ GeV,
$a^K_1(1GeV)=0.09$, $a^K_2(1GeV)=0.10$, $\delta_K^2=0.26$ GeV$^2$
and $\epsilon_K=0.7875$ and the upper edge is obtained by setting
$m_b=4.85$ GeV, $a^K_1(1GeV)=0.03$, $a^K_2(1GeV)=0.40$,
$\delta_\pi^2=0.14$ GeV$^2$ and $\epsilon_K=0.2625$. The main
uncertainties are caused by the value of $m_b$, $a^K_1$ and $a^K_2$,
and it can be found that all the $B\to K$ form factors shall
increase with the increment of $m_b$ and $a^K_2$, and decrease with
the increment of $a^K_1$. As for the LO results, we obtain
$[f^{T}_{B\to K}(0)/f^{+}_{B\to K}(0)]_{LO}\in [1.19,1.22]$; while
for the NLO results, we obtain $[f^{T}_{B\to K}(0)/f^{+}_{B\to
K}(0)]_{NLO} \in [1.09,1.15]$. Furthermore, we show the different
parts' contributions to the $B\to K$ form factors $f^{+,0,T}_{B\to
K}(q^2)$ in Fig.(\ref{fksep}), where all the parameters are taken to
be their center values. For $f^{+,0}_{B\to K}(q^2)$, it can be found
that the LO twist-2, the NLO twist-2 and the LO twist-4
contributions are positive, more specifically at $q^2=0$, they are
about $67\%$, $27\%$ and $6\%$ respectively. Even though the LO
twist-2 gives zero contribution to the sum of the form factor
$[f^{+}_{B\to K}+f^{-}_{B\to K}]$ but due to $SU_f(3)$-breaking
effect they shall give sizable contribution to $f^{0}_{B\to
K}(q^2)$, so $f^{0}_{B\to K}(q^2)$ is higher than the LO result
derived from Eq.(\ref{relation0a}) as shown in Fig.(\ref{fksep}).
For $f^{T}_{B\to K}(q^2)$, the LO twist-2, the NLO twist-2 give
positive contribution while the LO twist-4 gives negative
contribution, more specifically at $q^2=0$, they are about $70\%$,
$32\%$ and $-2\%$ respectively.

\section{Comparative studies of $f^{+,0,T}_{B\to\pi,\; K}(q^2)$
with other approaches in QCD LCSRs}

\subsection{A striking advantage of the present approach with the
chiral current}

The adopted chiral current approach has a striking advantage that
the twist-3 LC functions which are not known as well as the twist-2
light-cone functions are eliminated, and then it is considered to
provide results with less uncertainties. On the other hand, by using
the standard weak current in the correlator as shown by
Eqs.(\ref{cc1},\ref{cc2}), it has been pointed out that the twist-3
contributions can contribute $\sim 30-40\%$ to the total
contribution \cite{bkr}. So to obtain a more accurate result, one
has to calculate the above correlator by including one-loop
radiative corrections to both the twist-2 and the twist-3
contributions. Such a calculation together with the updated pion and
kaon twist-3 wave functions has been done by Ref.\cite{pball2}.

It may be interesting to do a comparison of their results with our
present ones so as to show whether these two treatments are
consistent with each other or not. For such purpose, we adopt the
following convenient form for the QCD sum rules obtained by
Ref.\cite{pball2}, which splits the $B\to P$ form factors into
contributions from different Gegenbauer moments:
\begin{equation}\label{sumruleapp}
F_{B\to P}^{+,0,T}(q^2)=f^{as}(q^2)+a^P_1(\mu_0)
f^{a^P_1}(q^2)+a^P_2(\mu_0) f^{a^P_2}(q^2)+a^P_4(\mu_0)
f^{a^P_4}(q^2),
\end{equation}
where $f^{as}$ contains the contributions to the form factor from
the asymptotic DA and all higher-twist effects from three-particle
quark-quark-gluon matrix elements, $f^{a^P_1,a^P_2,a^P_4}$ contains
the contribution from the higher Gegenbauer term of DA that is
proportional to $a^P_1$, $a^P_2$ and $a^P_4$ respectively. The
explicit expressions of $f^{as,a^P_1,a^P_2,a^P_4}$ for all the
mentioned form factors can be found in Table V and Table IX of
Ref.\cite{pball2}. And in doing the comparison, we take the same DA
moments for both methods.

\begin{figure}
\centering
\includegraphics[width=0.3\textwidth]{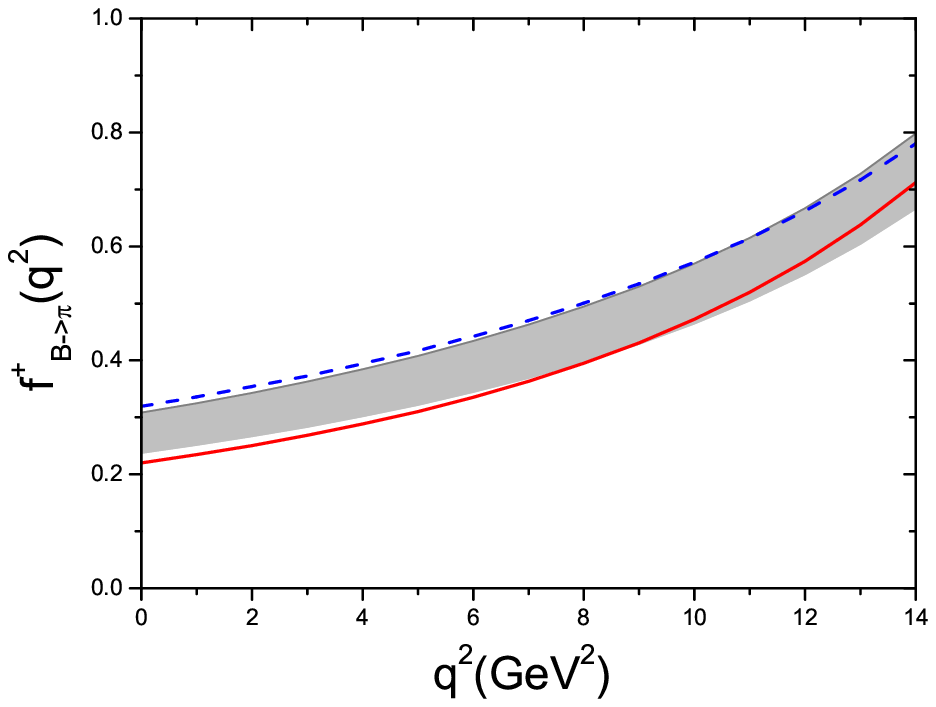}
\includegraphics[width=0.3\textwidth]{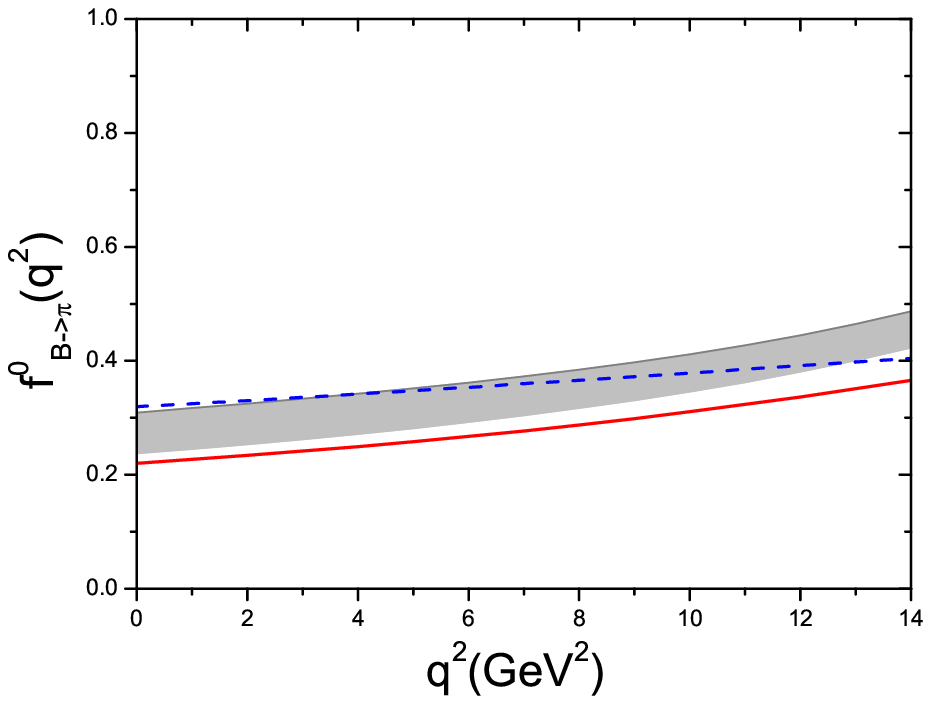}
\includegraphics[width=0.3\textwidth]{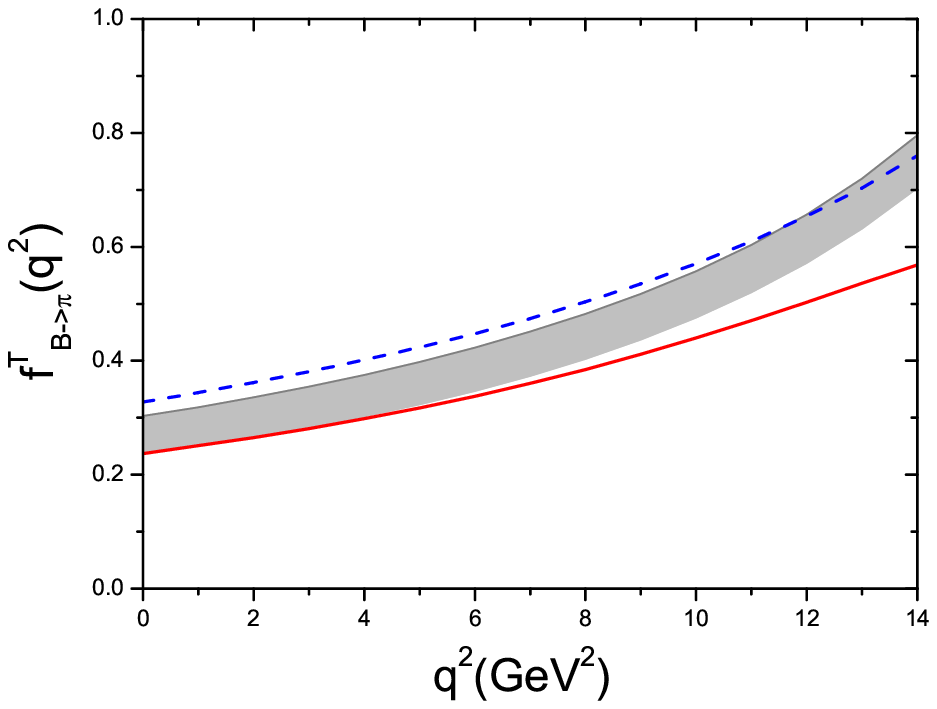}
\caption{$f_{B\to\pi}^{+,0,T}(q^2)$ with the allowed values of
$a^\pi_2$ and $a^\pi_4$ being correlated and given by the rhomboid
shown in Fig.\ref{a24pi}. The solid line is obtained with
$a^\pi_2(1GeV)=0.0$ and $a^\pi_4(1GeV)=0.0$ and the dashed line is
obtained with $a^\pi_2(1GeV)=0.23$ and $a^\pi_4(1GeV)=-0.030$, which
set the upper and the lower ranges of $f_{B\to\pi}^{+,0,T}(q^2)$
respectively. As a comparison, the shaded band shows the results of
Ref.\cite{pball2} together with its $12\%$ theoretical uncertainty.}
\label{comparepi}
\end{figure}

\begin{figure}
\centering
\includegraphics[width=0.3\textwidth]{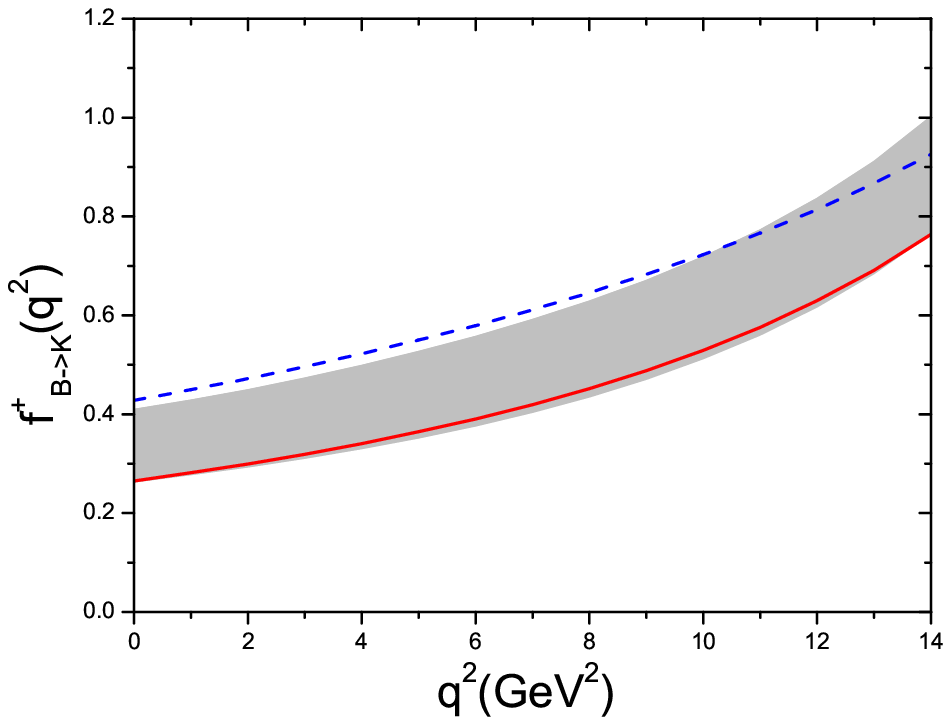}
\includegraphics[width=0.3\textwidth]{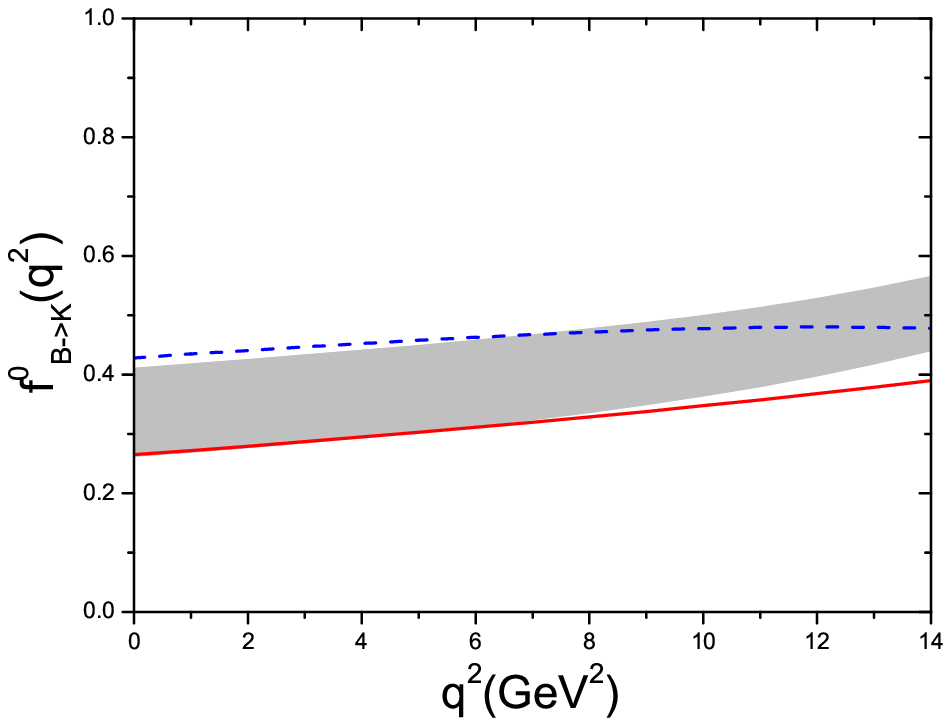}
\includegraphics[width=0.3\textwidth]{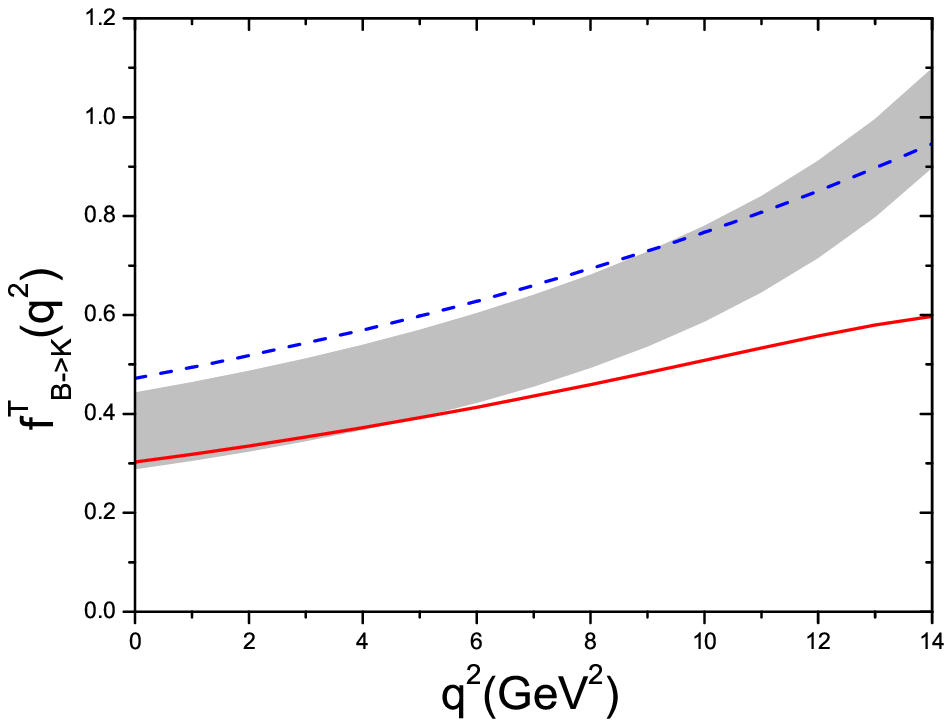}
\caption{$f_{B\to K}^{+,0,T}(q^2)$ for $a^K_1(1GeV)\in[0.03,0.09]$
and $a^K_2(1GeV)\in[0.10,0.40]$. The solid line is obtained with
$a^K_1(1GeV)=0.09$ and $a^K_2(1GeV)=0.10$ and the dashed line is
obtained with $a^K_1(1GeV)=0.03$ and $a^K_2(1GeV)=0.40$, which set
the upper and the lower ranges of $f_{B\to K}^{+,0,T}(q^2)$
respectively. As a comparison, the shaded band shows the results of
Ref.\cite{pball2} together with its $15\%$ theoretical uncertainty.}
\label{comparek}
\end{figure}

We show a comparison of our results of $F_{B\to\pi,K}^{+,0,T}(q^2)$
with those of Eq.(\ref{sumruleapp}) in
Figs.(\ref{comparepi},\ref{comparek}) respectively.
Fig.(\ref{comparepi}) shows $f_{B\to\pi}^{+,0,T}(q^2)$ with
$a^\pi_2$ and $a^\pi_4$ being correlated and given by the rhomboid
shown in Fig.(\ref{a24pi}), where the solid line is obtained with
$a^\pi_2(1GeV)=0.0$ and $a^\pi_4(1GeV)=0.0$ and the dashed line is
obtained with $a^\pi_2(1GeV)=0.23$ and $a^\pi_4(1GeV)=-0.030$, which
set the upper and the lower ranges of $f_{B\to\pi}^{+,0,T}(q^2)$
respectively. Fig.(\ref{comparek}) shows $f_{B\to K}^{+,0,T}(q^2)$
with $a^K_1(1GeV)\in[0.03,0.09]$ and $a^K_2(1GeV)\in[0.10,0.40]$,
where the solid line is obtained with $a^K_1(1GeV)=0.03$ and
$a^K_2(1GeV)=0.10$ and the dashed line is obtained with
$a^K_1(1GeV)=0.09$ and $a^K_2(1GeV)=0.40$, which set the upper and
the lower ranges of $f_{B\to K}^{+,0,T}(q^2)$ respectively. As a
comparison, the shaded bands in these figures show the results of
Eq.(\ref{sumruleapp}) within the same $a^K_1$ and $a^K_2$ region and
with their estimated $[12\%+3\%]$ theoretical uncertainty, where the
extra $3\%$ uncertainty is from $a^K_1$ uncertainty \cite{pball2}.

More explicitly, we show the comparison in detail:
\begin{itemize}

\item At the large recoil region $q^2=0$, Ref.\cite{pball2} gives
$f^{+,0}_{B\to\pi}(0)=0.258\pm 0.031$, $f^{T}_{B\to\pi}(0)=0.253\pm
0.028$, $f^{+,0}_{B\to K}(0)=0.304\pm 0.076$ and $f^{T}_{B\to K}(0)
=0.332\pm0.080$. It can be found that our results as shown by
Eqs.(\ref{ffbpi},\ref{ffbk}) are consistent with those of
Ref.\cite{pball2}, especially in the lower $q^2$ region.

\item With the increment of $q^2$, the form factors of
Ref.\cite{pball2} increase faster than ours. We can see this clearly
from the scalar and tensor form factors $f_{B\to\pi,K}^{0,T}(q^2)$.
These differences, especially in the larger $q^2$ region are mainly
caused by the treatment of the twist-3 contribution and by the
different treatment of the uncertainty. The twist-3 contribution can
affect the shape of the form factors. For example, as shown by
Fig.(\ref{fpisep}), the present obtained $f^{0}_{B\to\pi,K}(q^2)$
are close to the LO result derived from Eq.(\ref{relation0a}); while
Ref.\cite{pball2} gives a larger $f^{0}_{B\to\pi,K}(q^2)$ at higher
$q^2$ region due to the fact that the twist-3 contribution to
$[f^{+}_{B\to\pi,K}+f^{-}_{B\to\pi,K}]$ is dominant over the leading
twist contribution at large momentum transfer. In Ref.\cite{pball2}
the total uncertainty is obtained by adding up the uncertainties
caused by each parameter in quadrature; while at the present, we
vary the parameters within their possible regions and adopt the
minimum and the maximum ones as the uncertainty boundary. Moreover,
we have adopted a simple overall uncertainty $12\%$ or $15\%$ for
the form factors within all $q^2$ for the LCSRs of
Ref.\cite{pball2}, which in fact should be varied according to
different $q^2$, e.g. we have found that such uncertainty may be up
to $5\%$ for the mentioned form factors with the region of $q^2\in
[0,14]\; GeV^2$.

\item One may observe that in the lower $q^2$ region,
different from Ref.\cite{pball2} where $F^{B\to K}_{+,0,T}(q^2)$
increases with the increment of both $a^K_1$ and $a^K_2$, our
present predicted $F^{B\to K}_{+,0,T}(q^2)$ will increase with the
increment of $a^K_2$ but with the decrement of $a^K_1$. This
difference is caused by the fact that we adopt the pion and kaon DAs
derived from their wave functions to do our discussion, whose
parameters are determined by the combined effects of $a^K_1$ and
$a^K_2$; while in Ref.\cite{pball2}, $a^K_1$ and $a^K_2$ are varied
independently and then their contributions are changed separately.

\end{itemize}

\subsection{A comparison of the choosing of pole or $\overline{MS}$ $b$-quark mass }

Refs.\cite{duplan,duplan2} has argued to used $\overline{MS}$
$b$-quark mass instead of the pole quark mass. The $\overline{MS}$
$b$-quark running mass ($\bar{m}_b$) is related to the one-loop
$b$-quark pole mass ($m_b^*$) through the following well-known
relation:
\begin{eqnarray}\label{relat}
\bar{m}_b(\mu)& = & m_b^* \left \{ 1 + \frac{\alpha_S(\mu) C_F}{4
\pi} \left(-4 + 3 \ln \frac{m_b^{*2}}{\mu^2} \right) \right \}.
\end{eqnarray}
With the help of the relation (\ref{relat}), one can conveniently
transform the form factor expressions among these two choices of
$b$-quark mass. And one only need to be careful to use all the
parameters calculated under the same choice, e.g. the value of $f_b$
should be calculated by using the same currents in the correlator
and under the same choice of $b$-quark mass. By calculating the
ordinary correlators (\ref{cc1},\ref{cc2}) up to NLO and by varying
the $\overline{MS}$ $b$-quark mass within the region of
$\bar{m}_b(\bar{m}_b)=4.164\pm0.025$ GeV, Refs.\cite{duplan,duplan2}
obtain $f^{+,0}_{B\to\pi}(0)=0.26^{+0.04}_{-0.03}$,
$f^{T}_{B\to\pi}(0)=0.255\pm 0.035$, $f^{+,0}_{B\to
K}(0)=0.36^{+0.05}_{-0.04}$, $f^{T}_{B\to K}(0) =0.38\pm0.05$, and
\begin{equation}
\frac{f^{+,0}_{B\to K}(0)}{f^{+,0}_{B\to\pi}(0)}=
1.38^{+0.11}_{-0.10} \;,\;\; \frac{f^{T}_{B\to
K}(0)}{f^{T}_{B\to\pi}(0)}=1.49^{+0.18}_{-0.06} .
\end{equation}
These results are consistent with ours and also with those of
Ref.\cite{pball2} within reasonable errors, which is also calculated
by taking the pole quark mass. This shows that these two choices of
$b$-quark mass are equivalent to each other.

\subsection{Extrapolations of the LCSR results to higher $q^2$ region}

In order to allow a simple implementation of our results, we present
a parametrization that includes the main features of the analytical
properties of the form factors and is valid in the full physical
regime $0\leq q^2 \leq(m_B-m_P)^2$. Following the same argument of
Ref.\cite{pball2}, we fit the LCSR results to the following
parametrizations that are based on the procedure advocated by
Becirevic and Kaidalov \cite{BK}, where we take the LCSR results
with all the parameters taken to be their center values to do the
extrapolation, i.e. the $b$-quark one-loop pole mass $m_b=4.8$ GeV,
$a^\pi_2(1GeV)=0.115$, $a^\pi_4(1GeV)=-0.015$, $a^K_1(1GeV)=0.06$
and $a^K_2(1GeV)=0.25$. To measure the quality of the fit, we
introduce the parameter $\Delta$ that is defined as
\begin{equation}
\Delta = 100\,\max_{t}\, \left| \frac{f(t)-f^{\rm fit}(t)}{f(t)}
\right| \,, \quad t \in \left\{0,\frac{1}{2},\dots,\frac{23}{2},
12\right\}\,\text{GeV}^2,
\end{equation}
i.e.\ it gives, in per cent, the maximum deviation of the fitted
formfactors from the original LCSR result for $q^2<12\,{\rm GeV}^2$.
\begin{itemize}
\item for $f_{+,T}^\pi$:
\begin{equation}
f(q^2) = \frac{r_1}{1-q^2/(m_1^{\pi})^2} + \frac{r_2}{1-q^2/m_{\rm
fit}^2},
\end{equation}
where $m_1^\pi=5.325$ GeV \cite{pdg} is the mass of $B^{*}(1^-)$.
For $f_{+}^\pi$, the fit parameters are $r_1=0.7411$, $r_2=-0.4815$
and $m_{\rm fit}^2=40.01$ ${\rm GeV^2}$ for $\Delta \simeq 0.05$.
And for $f_{T}^\pi$, the fit parameters are $r_1=0.7742$,
$r_2=-0.4952$ and $m_{\rm fit}^2=34.71$ ${\rm GeV^2}$ for $\Delta
\simeq 0.9$.

\item for $f_{+,T}^{K}$:
\begin{equation}
f(q^2) = \frac{r_1}{1-q^2/(m_1^{K})^2} + \frac{r_2}{1-q^2/m_{\rm
fit}^2},
\end{equation}
where $m^K_1=5.413$ GeV \cite{pdg} is the mass of the
$B^{*}_{s}(1^-)$. For $f_{+}^K$, the fit parameters are
$r_1=0.8182$, $r_2=-0.4862$ and $m_{\rm fit}^2=41.61$ ${\rm GeV^2}$
for $\Delta \simeq 1.3$. And for $f_{T}^K$, the fit parameters are
$r_1=0.893$, $r_2=-0.5073$ and $m_{\rm fit}^2=33.13$ ${\rm GeV^2}$
for $\Delta \simeq 1.8$.

\item for $f^{\pi,K}_0$:
\begin{equation}
f_0(q^2) = \frac{r_2}{1-q^2/m_{\rm fit}^2} .
\end{equation}
For the case of pion, we obtain $r_2=0.2596$ and $m_{\rm
fit}^2=46.09$ ${\rm GeV^2}$ for $\Delta=0.07$. For the case of kaon,
we obtain $r_2=0.332$ and $m_{\rm fit}^2=61.64$ ${\rm GeV^2}$ for
$\Delta=1.3$.
\end{itemize}

\begin{figure}
\centering
\includegraphics[width=0.4\textwidth]{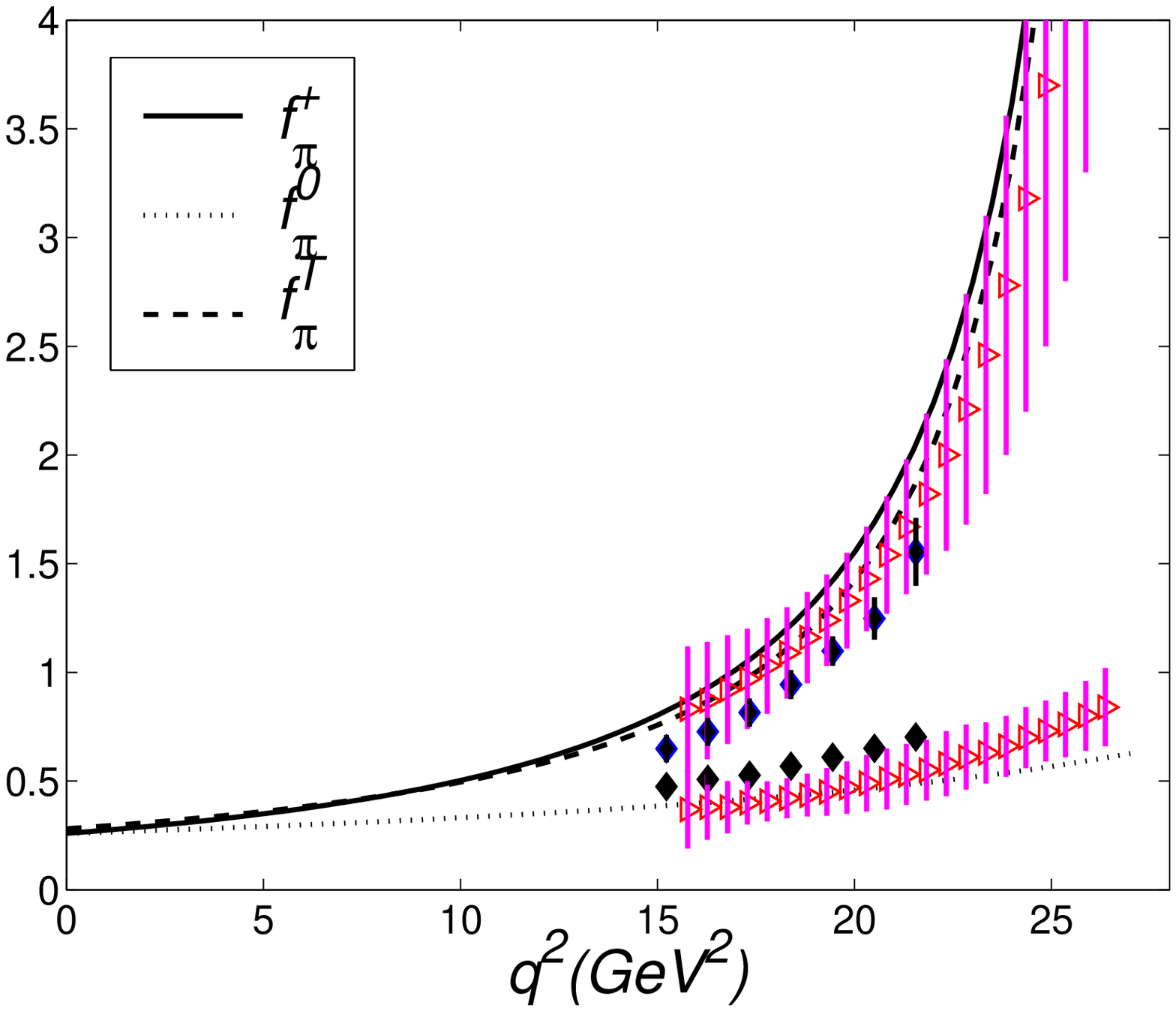}\hspace{1.0cm}
\includegraphics[width=0.4\textwidth]{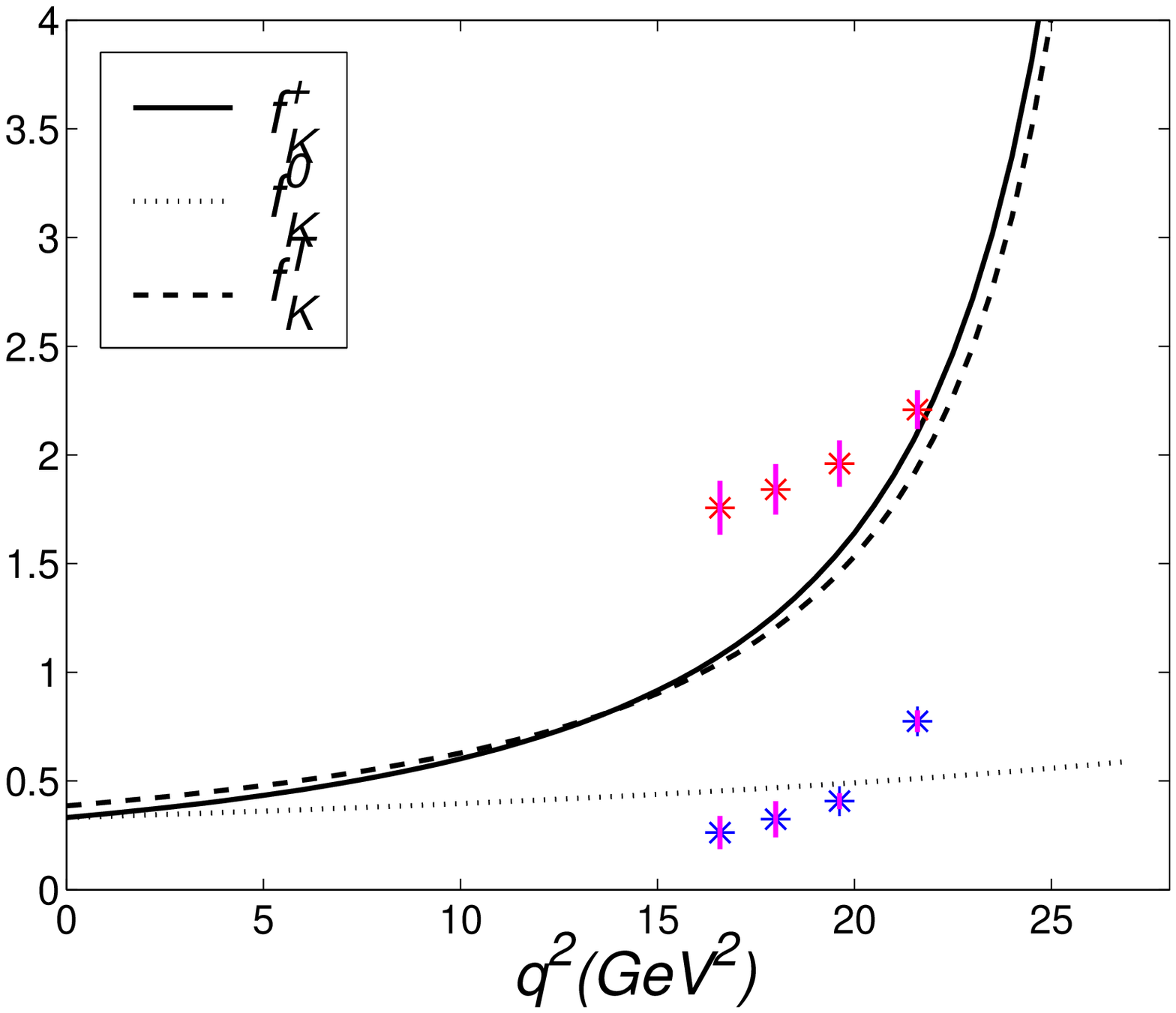}
\caption{Extrapolations of the LCSR Results of $B\to\pi$ and $K$
form factors for higher $q^2$. For comparison, Left diagram shows
the unquenched lattice QCD result \cite{latticebpi1} (Diamond) and
the quenched lattice QCD result \cite{JLQCD} (Triangle) with their
corresponding errors for the vector and scalar $B\to\pi$ form
factors; Right diagram shows the Lattice QCD results \cite{lattBK}
for the vector and scalar $B\to K$ form factors (Asterisk).}
\label{fitff}
\end{figure}

A comparison of the lattice QCD results can be found in
Fig.(\ref{fitff}). There are many Lattice results in the literature
for $B\to\pi$, e.g.
\cite{fermilab,JLQCD,latticebpi1,latticebpi2,abada,khadra,lattBK}
and etc., for convenience, we have taken the unquenched lattice QCD
result \cite{latticebpi1} and the quenched lattice QCD result
\cite{JLQCD} of $B\to\pi$ form factors. While for the $B\to K$ form
factors there is little lattice QCD results, and we adopt the
preliminary results derived by Ref.\cite{lattBK} to do our
discussion.

\section{Summary}

We have calculated all the $B\to P$ transition form factors
$F^{+,0,T}_{B\to P}(q^2)$ with chiral current in the LCSR up to NLO,
in which the most uncertain twist-3 contributions have been
eliminated naturally, with the b-quark pole mass that is universal.
The $SU_f(3)$-breaking effects in $B\to K$ form factors have been
carefully discussed and their values depend on the moment of the
kaon distribution amplitude, $a^K_2(1GeV)$. It is found that
$\frac{f^{+,0}_{B\to K}(0)}{f^{+,0}_{B\to\pi}(0)}=
1.28^{+0.06}_{-0.08}$ and $ \frac{f^{T}_{B\to
K}(0)}{f^{T}_{B\to\pi}(0)}=1.37^{+0.07}_{-0.02}$ for $a^K_1(1GeV)\in
[0.03, 0.09]$ and $a^K_2(1GeV)\in [0.10, 0.40]$. Based on the LCSR
with chiral current, we have made a comparative study on the
properties of transition form factors with those obtained in
literature \cite{pball2,duplan,duplan2}, in which the radiative
corrections on both the twist-2 and twist-3 parts should be treated
in equal footing. It has been found that the present results are
less uncertain under the same parameter regions to consider the
radiative corrections since the twist-3 contributions have been
eliminated naturally in the adopted method, so our results are
simpler and consistent with those in literature that have been
derived with the usual correlators. These form factors are important
ingredients in the analysis of semileptonic $B$ decays, especially
our present results may be helpful to clarify the present conditions
for the $B\to \eta^{(\prime)}(\ell^{-} \bar\nu_{\ell},\, \ell^{+}
\ell^{-})$ decays and then a better understanding of the $\eta$ and
$\eta'$ mixing \cite{whnew}.

\vspace{1cm} {\bf Acknowledgments}: This work was supported in part
by Natural Science Foundation Project of CQ CSTC under Grant
No.2008BB0298 and Natural Science Foundation of China under Grant
No.10805082 and Grant No.10475084, and by the grant from the Chinese
Academy of Engineering Physics under Grant No.2008T0401 and Grant
No.2008T0402. \\

\appendix

\section{Pion and kaon distribution amplitudes}

Generally, the pion and kaon twist-2 and twist-4 DAs can be written
in the following forms:
\begin{itemize}
\item twist-2 DAs:
\begin{equation}\label{phigen}
\varphi_{P}(u,\mu)=6u\bar{u}\Big[1 +{a^P_1}(\mu) C_1^{3/2}(2u-1) +
{a^P_2}(\mu) C_2^{3/2}(2u-1) + {a^P_4}(\mu) C_4^{3/2}(2u-1)+\cdots
\Big]\,,
\end{equation}
where $P$ stands for $\pi$ or $K$ respectively, $\cdots$ stands for
even higher Gegenbauer terms.

\item twist-4 DA's \cite{wk1}:
\begin{eqnarray}
\Psi_{4P}(\alpha_i) &=& 30 \alpha_3^2(\alpha_2-\alpha_1)\left[
h_{00}+h_{01}\alpha_3+\frac{1}{2}\,h_{10}(5\alpha_3-3)\right] , \nonumber\\
\widetilde{\Psi}_{4P}(\alpha_i) &=& -30 \alpha_3^2\left[
h_{00}(1-\alpha_3)+h_{01}\Big[\alpha_3(1-\alpha_3)
-6\alpha_1\alpha_2\Big]\right.\nonumber\\
&&\left. +h_{10}\Big[\alpha_3(1-\alpha_3)-\frac{3}{2}(\alpha_1^2
+\alpha_2^2)\Big]\right],\nonumber\\
\Phi_{4P}(\alpha_i) &=& 120 \alpha_1\alpha_2\alpha_3
\left[ a_{10} (\alpha_1-\alpha_2)\right],\nonumber\\
\widetilde{\Phi}_{4P}(\alpha_i) &=& 120 \alpha_1\alpha_2 \alpha_3
\left[ v_{00} + v_{10} (3\alpha_3-1)\right],
\end{eqnarray}
where
\begin{eqnarray}
h_{00} & = & v_{00} = -\frac{M_P^2}{3}\,\eta_{4P}=-\frac{\delta_P^2}{3},\nonumber \\
a_{10} & = & \frac{21M_P^2}{8}\eta_{4P}\omega_{4P} -\frac{9}{20}\,
a^P_2 M_P^2 =\delta^2_P \epsilon_P-\frac{9}{20} a^P_2 M_P^2,\nonumber\\
v_{10} & = & \frac{21M_P^2}{8} \eta_{4P} \omega_{4P}=\delta_P^2 \epsilon_P,\nonumber\\
h_{01} & = & \frac{7M_P^2}{4} \eta_{4P} \omega_{4P} -\frac{3}{20}
a^P_2 M_P^2=\frac{2}{3}\delta^2_P \epsilon_P - \frac{3}{20} a^P_2
M_P^2 \nonumber
\end{eqnarray}
and
\begin{displaymath}
h_{10} = \frac{7M_P^2}{2} \eta_{4P} \omega_{4P} + \frac{3}{20} a^P_2
M_P^2=\frac{4}{3}\delta^2_P \epsilon_P +\frac{3}{20} a^P_2 M_P^2 ,
\end{displaymath}
with $\eta_{4P}=\delta^2_P/M_P^2$, $\omega_{4P}=8\epsilon_P /21$.
Taking the leading meson-mass effect into consideration, the
remaining two-particle DA's of twist 4 can be written as \cite{wk1}:
\begin{eqnarray}
\phi_{4P}(u) &=& \frac{4u\bar{u} }{3}\Big\{ -5u\bar{u}\left[
30h_{00} + 4h_{01}( 3 + u\bar{u})  +5h_{10}( -3 + 2u\bar{u}) \right]
\nonumber\\
&& +2a_{10}[ 6 + u\bar{u}( 9 + 40u\bar{u})]\Big\}+
8a_{10}\Big\{2u^3(10 -15u + 6u^2)\ln u \nonumber \\
&& +2\bar{u}^3(10 -15\bar u + 6\bar u^2)\ln \bar u \Big \}\,,\\
\psi_{4P}(u)&=&5\Big[ -4h_{00} - 2h_{01} + h_{10} + 4\left( -4a_{10}
+ 6h_{00} + 4h_{01} + h_{10} \right) u  \nonumber\\
&& -6\left(4h_{00} + 6h_{01} +9h_{10}-16a_{10} \right) u^2 +
20\left(2h_{01} + 5h_{10}-8a_{10} \right) u^3 \nonumber\\
&&+10\left( 8a_{10} - 2h_{01} - 5h_{10} \right) u^4 \Big]\,.
\end{eqnarray}
Setting $M_P \to 0$ and $M_P \to M_K$, one can obtain the pionic and
the kaonic twist-4 DAs respectively.
\end{itemize}

\end{document}